# Design, Implementation, and Performance of the Primary Reflector for SALTUS

**Jonathan W. Arenberg,[a,*] Leon K. Harding,[a] Bob Chang,[a] Steve Kuehn,[a] Dave Oberg,[a] Michaela N. Villarreal,[a] , Arthur L. Palisoc,[b] Christopher Walker,[c] Daewook Kim,[c] Zach Lung,[d] Dave Lung[d]**

[a]Northrop Grumman, Dulles, Virginia 20166, United States
[b]L'Garde, Inc., 15181 Woodlawn Ave, Tustin, California 92780-6487, United States
[c]Department of Astronomy, University of Arizona 933 N Cherry Ave., Tucson, Arizona 85721, United States
[d]DA2 Ventures, 3970 Ridgeline Drive, Timnath, Colorado 80547, United States

**Abstract**. The Single Aperture Large Telescope for Universe Studies (SALTUS) is a mission concept for a far-infrared observatory developed under the recent Astrophysics Probe Explorer opportunity from NASA. The enabling element of the program is a 14 m diameter inflatable primary mirror, M1. Due to its importance to SALTUS and potentially other space observatories, this paper focuses entirely on M1. We present a historical overview of inflatable systems, illustrating that M1 is the logical next step in the evolution of such systems. The process of design and manufacture is addressed. We examine how M1 performs in its environment in terms of operating temperature, interaction with the solar wind, and shape change due to non-penetrating particles. We investigate the longevity of the inflatant in detail and show it meets mission lifetime requirements with ample margin and discuss the development and testing to realize the flight M1.

*Jonathan W. Arenberg, E-mail: jon.arenberg@ngc.com

## 1 Introduction

The Single Aperture Large Telescope for Universe Studies (SALTUS, Latin for '*leap*') is a mission concept developed for the ongoing NASA Astrophysics Probe Explorer (APEX) opportunity. SALTUS is a far-infrared (far-IR) space observatory operating from a halo orbit around L2. SALTUS is capable of observing the first galaxies, protoplanetary disks at various evolutionary stages, and a wide variety of solar system objects, utilizing the large photon collecting power and angular resolution afforded by its 14-m diameter unobscured clear aperture covering a wavelength range from 34 to 659 µm. The science program for this observatory is described at length in the other contributions in this special issue. [1]

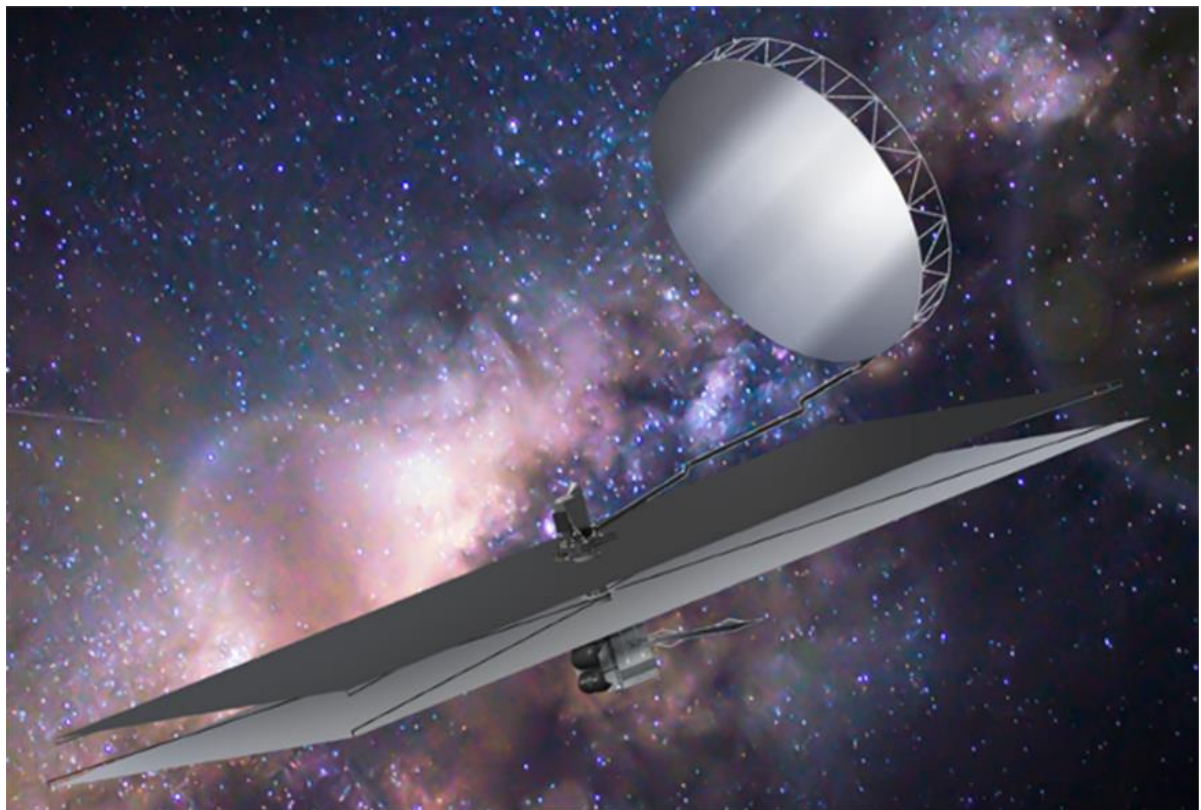

**Fig. 1** 3D rendered image of the SALTUS observatory, featuring an inflatable 14-m diameter off-axis primary mirror technology and its sunshield for radiative cooling below 45K (image credit University of Arizona).

The key enabling feature of the SALTUS architecture is its large aperture inflatable primary reflector known as M1, which is connected to the spacecraft via a deployable ~17-m single boom. [2] The lenticular reflector is shown in the context of the SALTUS mission in the sketch in Fig. 1and is constructed of 0.5 mil (6.75 µm) thick polyimide film. This paper is dedicated to addressing the most obvious questions surrounding such a large non-traditional optic. How is such a mirror designed and constructed? How can such a system maintain performance in the hostile environment of space? Furthermore, if such performance is established, how long can it be maintained? What are the plans for development and test?

The possibility of inflatable systems for applications in space is almost as old as the space age itself. In fact, one of the early demonstrations of the relevance of advanced space technology came with Echo-star and the first demonstration of advanced global communication.[3] Inflatable systems have been studied for various applications; structures, reflectors, and even habitats. [4,5,6,7,8] These activities, and in particular those concerned with making reflectors was very active and well-funded in the 1990s, with much of the art summarized in a book edited by Jenkins. [9] The death of research following 9/11 has left the research of inflatable systems largely stagnant. The literature through the 2000's on inflatable optic systems focused on the inflatable technology itself, and not its integration into a full system.

The fundamental promise of a very large aperture inflatable is the ability to enable unprecedented science via very large collecting areas; for low mass, volume, cost, and schedule. We are participants in a recent APEX proposal, SALTUS, based on this fundamental promise. The purpose of this paper is to explore and document our work in redeeming the promise of inflatable telescopes and building a mission relevant architecture around it. It is our intention to present our analysis in as general a manner as possible, using the work we did for SALTUS to build upon the prior conceptual design of OASIS, an Astrophysics Medium Explorer proposal (MIDEX). [40]

Inflatable systems have been part of the aerospace milieu since 1960, with Project Echo [3], and other early experiments. [9,10] There are also many examples of inflatable systems being considered as the basis of space based optical systems. [4,6,8,11,12] To underscore the idea that inflatable optical systems are hardly a novel concept, consider the results of literature search for inflatable reflectors and antennas for space, shown in Fig. 2. The figure illustrates that this basic architecture has been investigated for over forty years. Reported activity and technical development of inflatable membrane telescopes peaked around the year 2000, largely under US government funding. Following the peak a lull has existed until our current efforts which if included in Fig. 2, would be off scale. SALTUS seeks to build on significant development work and prior investments to evolve inflatable design architecture for astronomical space-based applications. A NASA experiment in the late 1990's, the Inflatable Antenna Experiment, a 14-m diameter parabolic reflector deployed from the space shuttle on Mission STS-77, has direct legacy to SALTUS. [13]

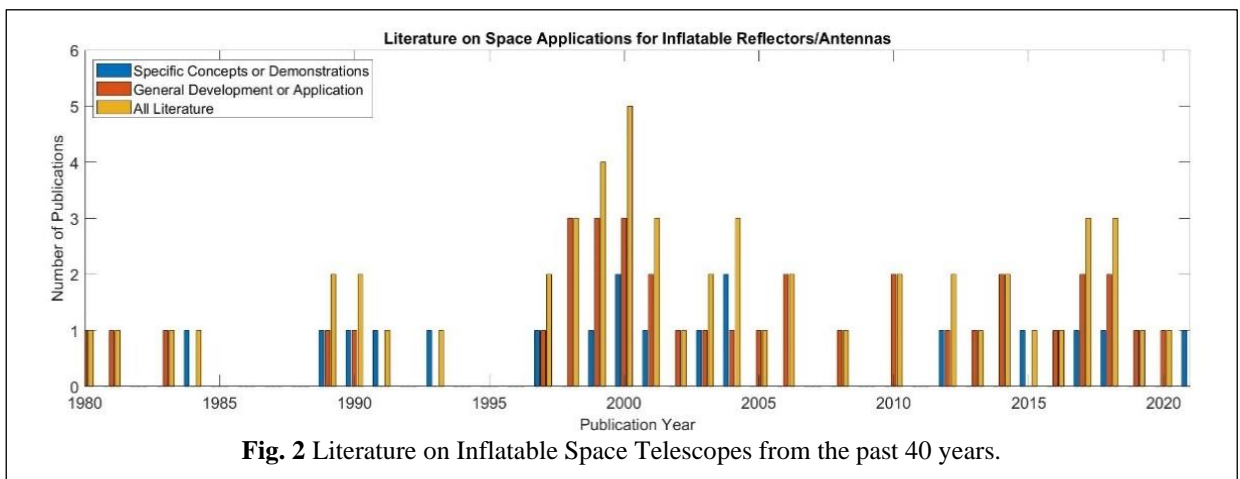

**Fig. 2** Literature on Inflatable Space Telescopes from the past 40 years.

Due to their lighter weight and more volumetrically efficient systems, inflatable optics are the natural next step in the evolution of space optics. This trend is driven by a fundamental truth of the physics of reflective optics; light from one's cosmic target only interacts with the reflective coating. Anthropomorphizing a bit, the cosmic signal only "knows' the coating layers and its properties, what holds that surface in place doesn't affect the optical performance. What matters is configuring the reflective surface in the right shape and at the right place in a system and maintaining performance. The inflatable system studied herein offers the possibility of creating and maintaining a very large aperture optical system, at low mass and volume, thereby enabling scientific discovery previously unimaginable. [1] This kind of promise seems almost too good to be true. To convince ourselves that this is not "21st century optical snake oil" we have methodically analyzed many of the key questions associated with the realization of a large space aperture based on inflatable optics.

A completely comprehensive treatment of the architecture of inflatable systems is well beyond the scope of a journal paper like the one you are reading. We focus on those elements of the architecture that we believe show inflatable systems are a viable architectural option for various mission types. The process of modeling an inflatable optic and the resulting optical design are separate reports [7,14,15]. For a system as large as SALTUS, we need to show that it is possible to measure the primary aperture and to do so without recourse to a vacuum chamber. [16] More details on other aspects of the SALTUS observatory architecture can be found in the paper in this special issue. [2]

It is also necessary to adequately understand and define the effects and conditions that lead to degradation of system performance. The design process adds features and functions to the system that mitigate the degradations from interactions with the environment to perform in a way that meets mission needs. SALTUS has yet to be built, so our argument for the viability of a very large aperture inflatable is by analysis, modeling, and demonstration. This establishes an engineering basis from which technical design challenges can be identified, addressed and mitigated.

Our discussion begins with a description of the pressure control system (PCS). A key part of the analysis of M1 is the application of the ideal gas law (IGL) and the kinetic theory of gasses (KT) to calculate the proper inflation and potential gas losses due to various causes. In order to evaluate potential impacts to the performance of an environmental stress, we compare the impact of that stress to the level derived for optical performance, denoted $\delta P$. If the impact of the environmental stress is much less than $\delta P$, it can be neglected. We will use a conservative analysis to determine if an effect is small, so we can have confidence in neglecting its impact at this stage of analysis. If an impact is found   not to be small compared with $\delta P$, then the M1 design must respond. This response can be either a design feature to mitigate the impact or some other process to reduce the risk of impact. This process has been applied in the analysis of other high-performance systems. [17]

The primary mirror system (PMS) consists of M1 and the surrounding truss that holds the reflector in tension and serves as the primary mechanical interface between M1 and the deployable boom. Both the boom and truss contain mechanisms to adjust and maintain alignment.

M1 is an off-axis parabolic-shaped inflatable formed by edge bonding two elliptical, 0.5mil (12.67 µm) thick, diameter polymer films. M1 has an optical clear aperture equivalent to a 14 m diameter, the physical size of M1 is larger, ~15 m diameter. The films are constructed from trapezoidal gore segments that are thermally formed on an aluminum mandrel to the proper figure, and edge bonded together in such a way that, when inflated, a seamless, off-axis parabolic shape is produced. M1 is filled with Far IR spectrally inert helium (He) and inflated to the design pressure, $P_0$ of ~5.1Pa. When operating space environment, these pressures produce a wrinkle-free surface. [15,18] The inflation pressure is actively controlled to ensure the correct figure is maintained during science observations over the mission lifetime.

Both the design and verification of M1 will be analytically based and validated on a subscale 8 meter and full flight size representative engineering models. To demonstrate that metrological data is sufficiently accurate to validate a finite element model (FEM), the SALTUS team carried out a series of measurements on an existing one-meter diameter test article in a chamber at the Northrop Grumman Space Park Facility. [7,15,16] This experimental series involved inflation with various gases at atmospheric pressure and vacuum, to quantify gravitational and buoyant effects. Test measurements and analytical models indicate the shape difference in M1 between being on-ground and on-orbit due to gravitational effects will be only in low orders. The low-order shape errors of M1, and the induced aberrations like coma, can be simply compensated by the rigid body motion of M1 with respect to the corrector unit. [14] This motion is facilitated by actuators. [2]

The AstroMesh™ reflector was developed in 1990 and has been refined over the last 30 years with thirteen successful mission deployments. [2, 18] A sketch of M1 installed in a deployed AstroMesh™ is shown in Fig. 3. The deployed AstroMesh™ reflector structural integrity comes from three components: the perimeter truss, the net consisting of kevlar webs, and the tension ties. The net is necessary to give the deployed truss radial stiffness by reacting tension and compression loading in the webs. Without the net the radial stiffness of the truss depends on the bending stiffness of the truss members. The tension ties preload the net so it can react compression loading.

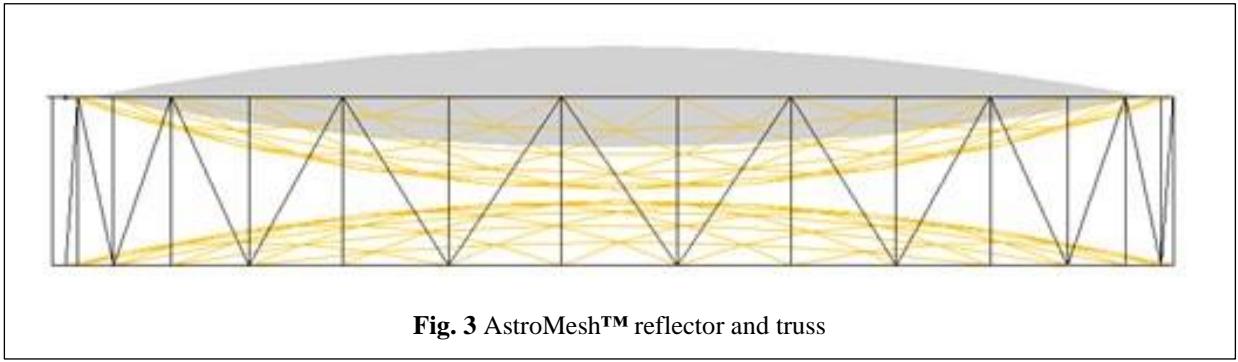

**Fig. 3** AstroMesh™ reflector and truss

## 2   Design and Manufacturing of M1 and Inflation Control System (ICS)

This section presents the methods, processes and heritage of the tools and processes used to design and manufacture M1 for the SALTUS mission.

### 2.1 Design Process

In order to achieve the M1's desired parabolic shape, we first have to solve the Inverse Problem: given the final desired surface shape, inflation pressure, material and geometric properties of the membrane, what must be the initial uninflated shape so that on inflation to the design pressure, the initial shape configuration transforms to the desired surface of revolution. The equations were derived and coded into a computer program FLATE, that outputs the initial un-stressed shape of the membrane gores. [20] The use of seam tapes to join the pie shaped gores of the parabolic membrane was also considered.

Fig. 4, panel a illustrates the Inverse Problem and

Fig. **4**, panel b shows how membrane gores are joined together to form the parabolic reflector. A canopy (as shown in

Fig. 4, panel c) is used to create a lenticular volume to contain the inflation pressure.

Utilizing the optics approach as described by Kim, et al. [14] millimeter level surface figure deviations across an aperture can be corrected to provide diffraction limited performance in the far-IR. Surface accuracies on the order of 0.5 mm to 1mm RMS have been measured on reflectors up to 3m in diameter. The 14 m diameter Inflatable Antenna Experiment (IAE) reflector had a measured accuracy of 2 mm RMS overall, with one 3m section having an error of 0.365 mm rms.[13] With the most recent fabrication method using very precise and accurate proprietary tooling and mandrel design, we expect surface accuracies of less than 1mm RMS for diameters up to 30 meters. We recently built and measured a 3m diameter, F/D=1 inflatable reflector to have a

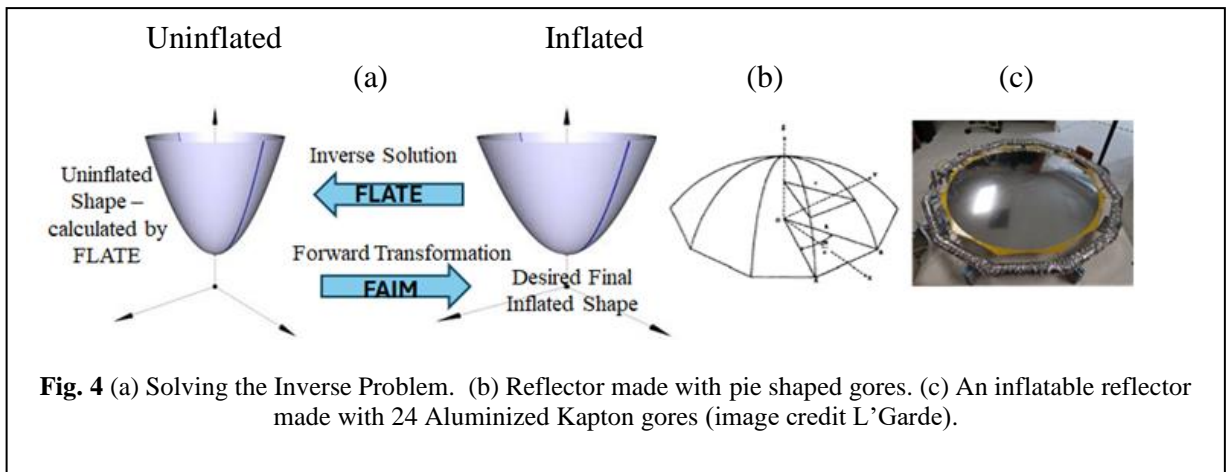

**Fig. 4** (a) Solving the Inverse Problem. (b) Reflector made with pie shaped gores. (c) An inflatable reflector made with 24 Aluminized Kapton gores (image credit L'Garde).

surface accuracy of ε=0.65 mm RMS. This test article was built on existing mandrels. If it was built using the more precise fabrication mandrels, the surface accuracy would be higher (smaller ε).

### 2.1.1 Preliminary Finite Element Analysis of SALTUS

Using the FAIM finite element code, we next present the preliminary shape accuracy analysis of the SALTUS 14-meter major diameter off-axis M1.[20] Fig. 6 shows the finite element model (FEM) mesh. We note at this point that the FAIM code used for the analyses has been validated against known analytical solutions as well as experimental results and showed excellent agreement. [15,19] FAIM is a geometrically nonlinear finite element analyzer with membranes and tension-only cables in its element library. It can accept loadings of internal and external pressure, nodal forces, element forces, nodal/element temperatures, body acceleration, and pre-stress. Boundary condition inputs include zero and non-zero displacements, spring and skew boundary conditions where nodes are restrained to only move along a plane not parallel to the global x-y, y-z, and x-z planes.

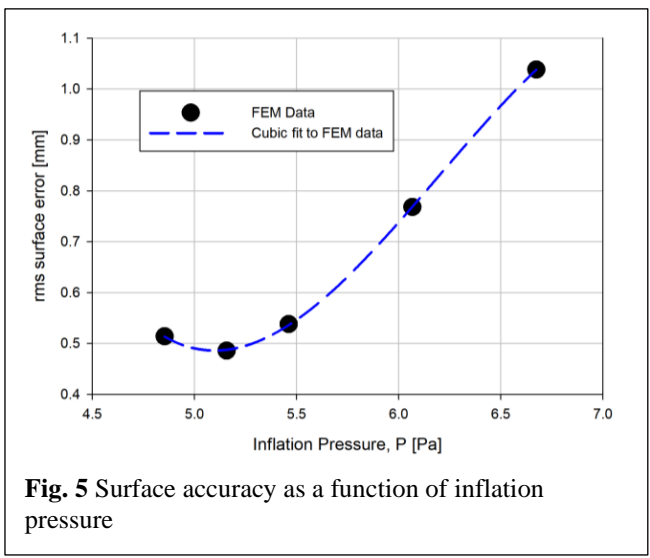

**Fig. 5** Surface accuracy as a function of inflation pressure

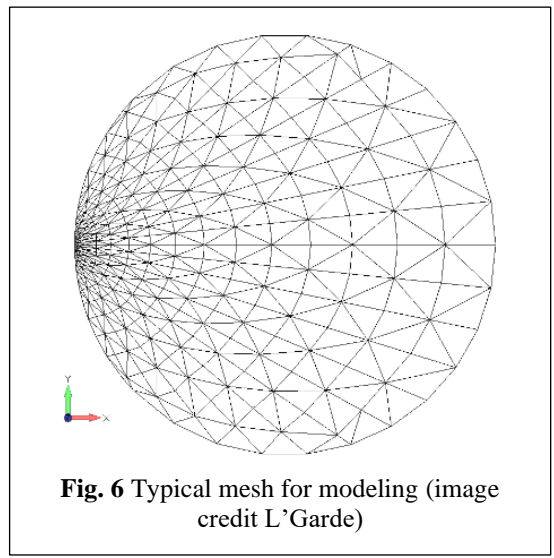

**Fig. 6** Typical mesh for modeling (image credit L'Garde)

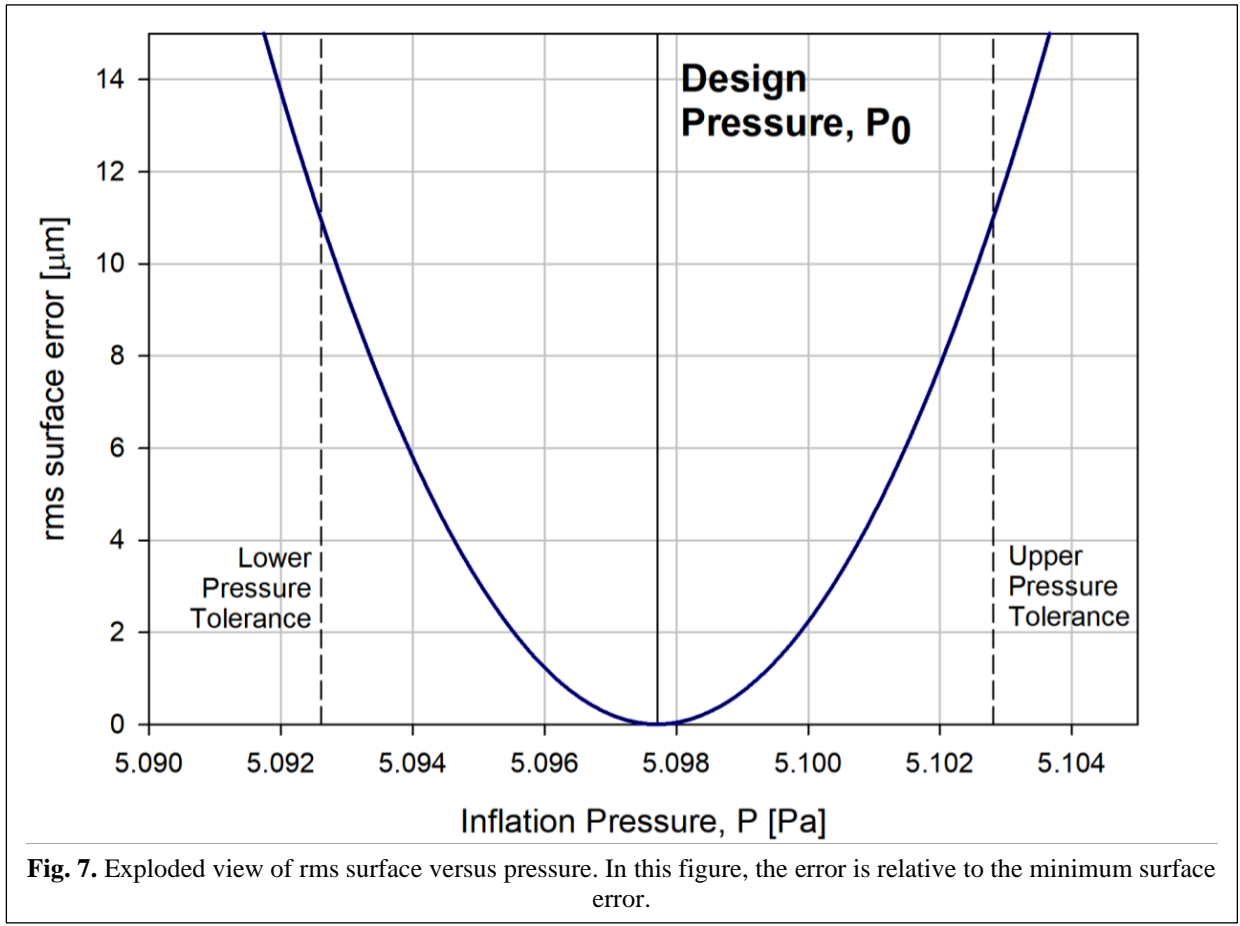

**Fig. 7.** Exploded view of rms surface versus pressure. In this figure, the error is relative to the minimum surface error.

The SALTUS uninflated seamless shape was first calculated using the FLATE code and was used to generate the geometry of the FEM. The inflation pressure was chosen to result in a film stress of approximately 6.895 MPa (1,000 psi). The boundary condition at the perimeter of the reflector consists of displacements of the outer perimeter nodes to their final inflated x-y-z location. Cable elements were used to simulate the effect of the seam tapes. The results of the FAIM runs are summarized in Fig. 5. Fig. 7 shows a zoomed in view near the minimum, showing the increase in rms error from the minimum, ~5.019 Pa, $P_0$, the nominal *designed-for* inflation pressure. The solid dots along the trace in Fig. 7 represent our initial inflation tolerance, $0.001P_0$ and shows that this error in inflation is equivalent to an increased surface error of 11 µm rms.

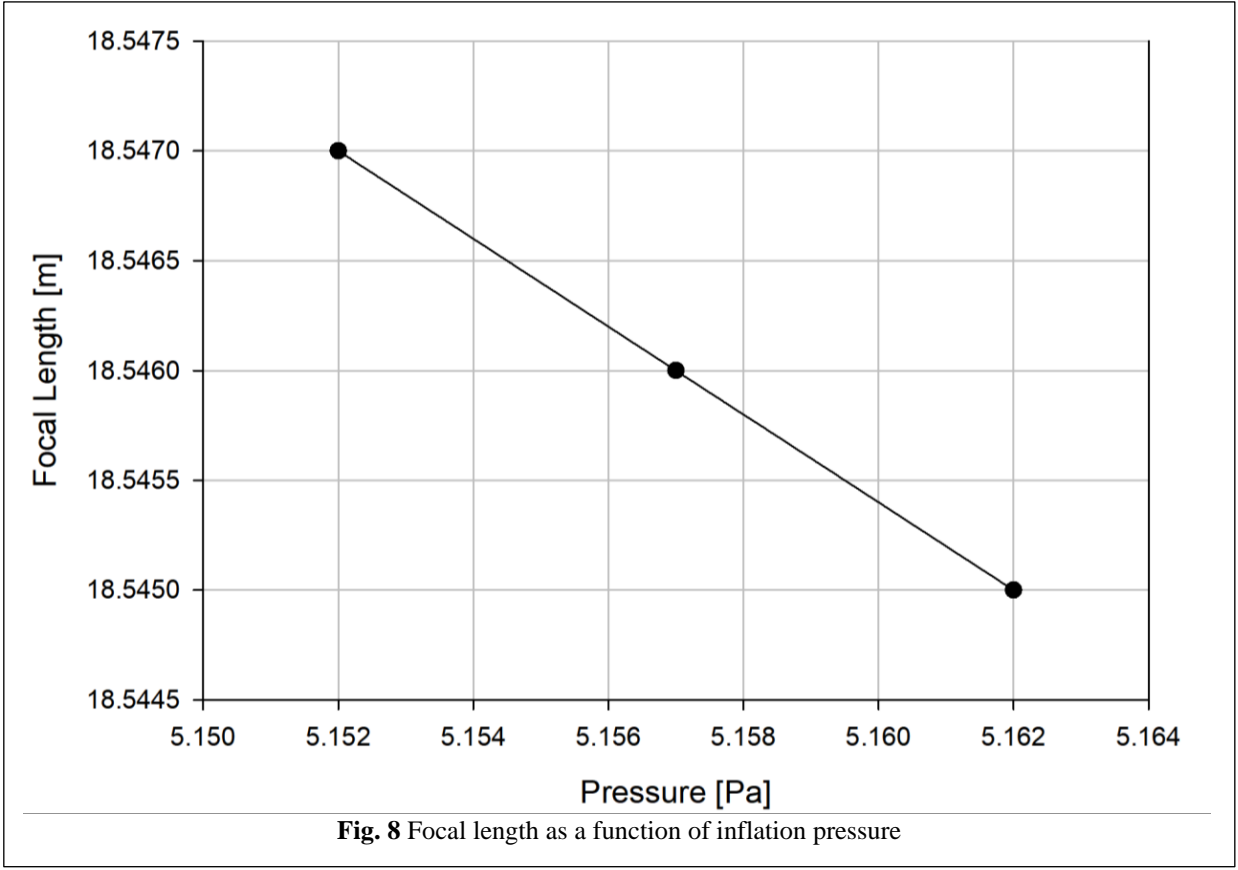

**Fig. 8** Focal length as a function of inflation pressure

Fig. 8 shows the change in focal length with M1 pressure.

FAIM serves as the crucial simulator for predicting the optical behavior of inflatable optics. Due to the dynamic nature of inflatable optics, anticipating fluctuations in surface variation becomes pivotal for evaluating and designing system performance. FAIM has demonstrated its capability to accurately compute varying surface errors both spatially and in terms of height. This analysis proves pivotal in crafting adaptable optics and system designs aimed at preserving collimation and ensuring fine spot diagrams. [19]

### 2.1.2 Tolerances and Shape Budget

The sources of inflatable reflector surface errors come from manufacturing and environmental factors. These include tolerances in (a) gore shape, (b) mounting to the perimeter truss, (c) material properties variation over the membrane continuum (modulus, coefficient of thermal expansion (CTE) , thickness, etc.), (d) thermal effects on orbit, and (e) material creep under heightened stress and temperature. The use of thermo-forming using highly accurate 3D gore mandrel assemblies will essentially reduce the gore shape error to its minimum dictated solely by the precision and accuracy of the mandrels.  The reduction of errors due to mating with the perimeter truss can be reduced by using highly accurate fabrication support equipment. SALTUS utilizes a sunshield and has a limited field of regard (FOR). The range of expected temperature variation over the FOR is less than 1K (hot to cold) and is discussed in Section 3.4.3. Moreover, the operating temperature of the SALTUS inflatable is ~37 K, rendering the effect of material creep to a minimum. [20]

The nominal membrane film stress on the inflatable SALTUS reflector is 1,000 psi, which makes M1 unwrinkled and therefore at least piecewise a smooth and continuous surface. We have also carried out measurements of the roughness of likely membrane materials for the suitability as specular reflectors, aka good mirrors. [21] In refence [21], we find that measured roughness of a typical membrane with coating is ~0.05 µm rms. Using the conventional approximation to total integrated scatter (TIS) from a surface at a reference wavelength of 30 µm is, approximately 0.00044. Thus, the scattered energy is a very small fraction compared to the specular reflection energy, which is 1-0.00044. Thus, we can confidently claim the M1 surface is specular, moreover this result has been confirmed experimentally many times. [16]

### 2.2 Optical Prescription and Performance

The on-orbit performance of M1 is predicted by using a validated FEM. The model validation comes from measuring the surface shape of the inflatable reflector on the ground as a function of inflation pressure, temperature and temperature variation over the surface, and boundary conditions. As discussed above, we will use our analytical tools designed specifically for membranes to guide the design (i.e., where to concentrate the effort in acheiving an accurate surface, should one be focusing on reducing a particular source of error, etc.). The results of the planned sensitivity analysis will guide the design to get optimal performance. As mentioned above, FAIM has been shown to accurately predict the optical shape and its results are used in an optical simulation. [14]

The manufacturing process and associated error budgets will be validated in our planned development articles of 8m diameter and a full-scale unit, see Section 4.

*2.3 Manufacturing*

The SALTUS M1 builds on the legacy of the 14-m diameter Inflatable Antenna L'Garde built for NASA in the mid-1990s. [13] It was flown off the Space Shuttle Mission STS-77 under the Inflatable Antenna Experiment (IAE). The geometric similarities include (a) the same offset geometry, (b) the same aperture major diameter of 14-m, and (c) similar structural construction, namely a symmetric canopy-reflector assembly consisting of identical gore shapes made of a thin polymeric material.

The main difference between the IAE and SALTUS are (a) the inflatant gas, (b) material used for the reflector-canopy, (c) a rigid truss, and most significantly, (d) the use of thermo-forming to achieve gores with the requisite double curvature, putting it closer to the target paraboloid

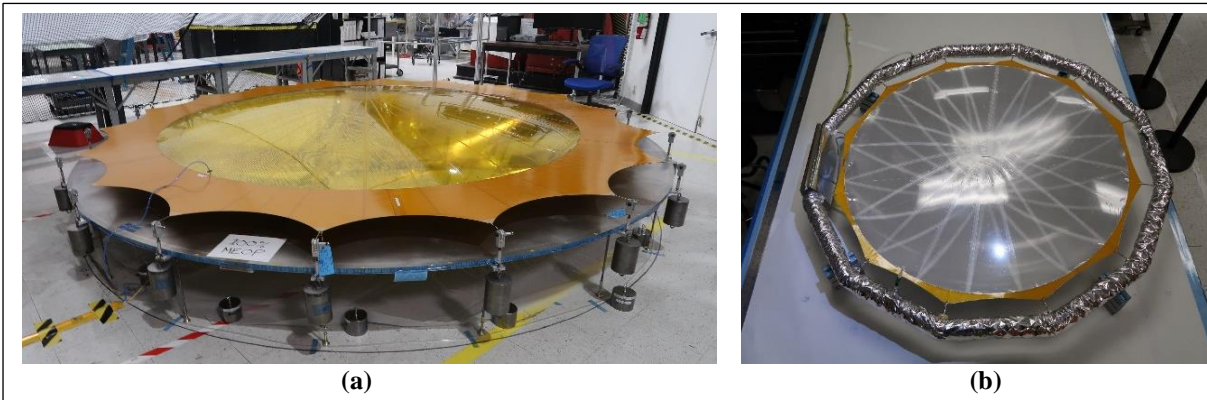

**Fig. 9** (a) 3m diameter inflatable lenticular showing its scalloped perimeter doubler edge support. (b) Fully inflated 1m diameter lenticular with inflatable torus support, diameter inflatable solar concentrator built by L'Garde for TransAstra Corporation in support of their *in-situ* resource utilization activities. (image credit L'Garde).

geometry. The IAE used nitrogen gas and the SALTUS inflatable reflector will be pressurized with helium (He) gas. For the lenticular material, whereas the IAE used 0.25 mil clear Mylar for the canopy and 0.25 mil aluminized Mylar for the reflector, the current SALTUS baseline call for clear 0.5 mil thick polyethylene and 0.5 mil thick aluminized polyimide for the canopy and reflector, respectively. Polyethylene was chosen because it has high transmittance at the wavelength of interest for the science and is tolerant to the environment. [47,48] These material choices will be studied and refined further as we proceed with development. The flight material membrane must meet a wide range of requirements that are yet to be fully defined, micrometeoroid impact robustness, electrical conductivity, coefficient of moisture, and thermal expansion, thickness uniformity, radiation, and UV damage resistance among others. Detailing these requirements and conducting the qualification plan will be activities done early in the program, very similar to the TRL-6 program carried out on the sunshield layers for the Webb Telescope at Northrop Grumman in the middle 2000's. [22]

As discussed in the previous section, the attainment of an accurate reflector surface starts with accurately shaped gores, the shape of which is derived from the calculation of the initial uninflated shape. Even with accurately and precisely cut gores, the desired high accuracy will not be there if the gores cannot be seamed together with as low an assembly error as possible. The seaming of all gores for both the canopy and reflector is followed by measurement validation of the just-built lenticular prior to bonding the perimeter doubler around the lenticular. An example of the (scalloped) perimeter edge support for an inflatable lenticular is shown in Fig. 9.

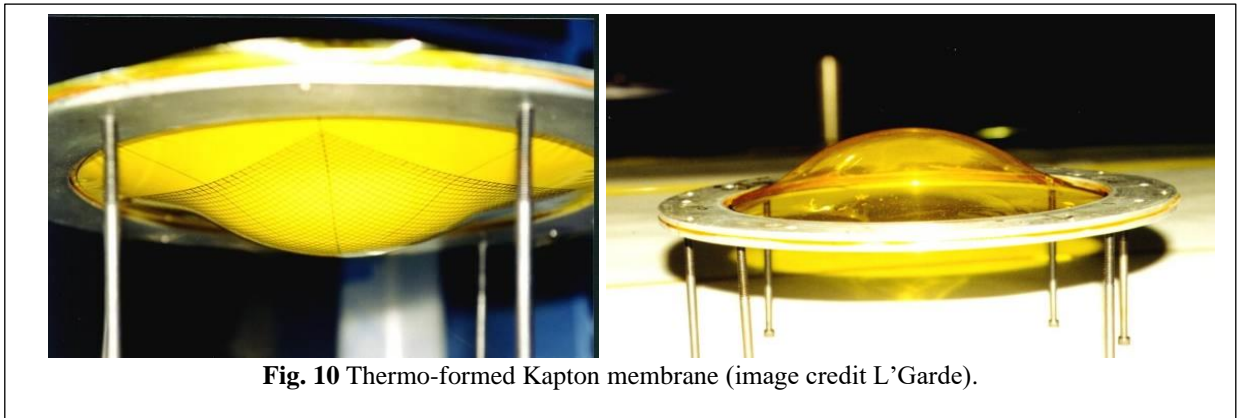

**Fig. 10** Thermo-formed Kapton membrane (image credit L'Garde).

### 2.3.1 Assembly, Integration and Measurement

Each gore of the 14-m diameter SALTUS M1 will be thermo-formed using a proprietary process. It uses a highly accurate 3D gore mandrel assembly.  An example of a thermo-formed Kapton membrane is shown in Fig. 10. [23] It is noted here that the curvature of the membrane shown in Fig. 10 is more severe than the curvature of the 14-m diameter SALTUS M1.

Fig. 11 shows the gores of the 3m diameter. These gores were made from flat gore segments (not thermo-formed) the surface figure accuracy obtained for the fully assembled 3m on axis reflector is *ε = 0.65 mm RMS,* which shows the robustness of the L'Garde code used. The use of a thermo-formed initial uninflated configuration for SALTUS will only enable higher surface accuracy.

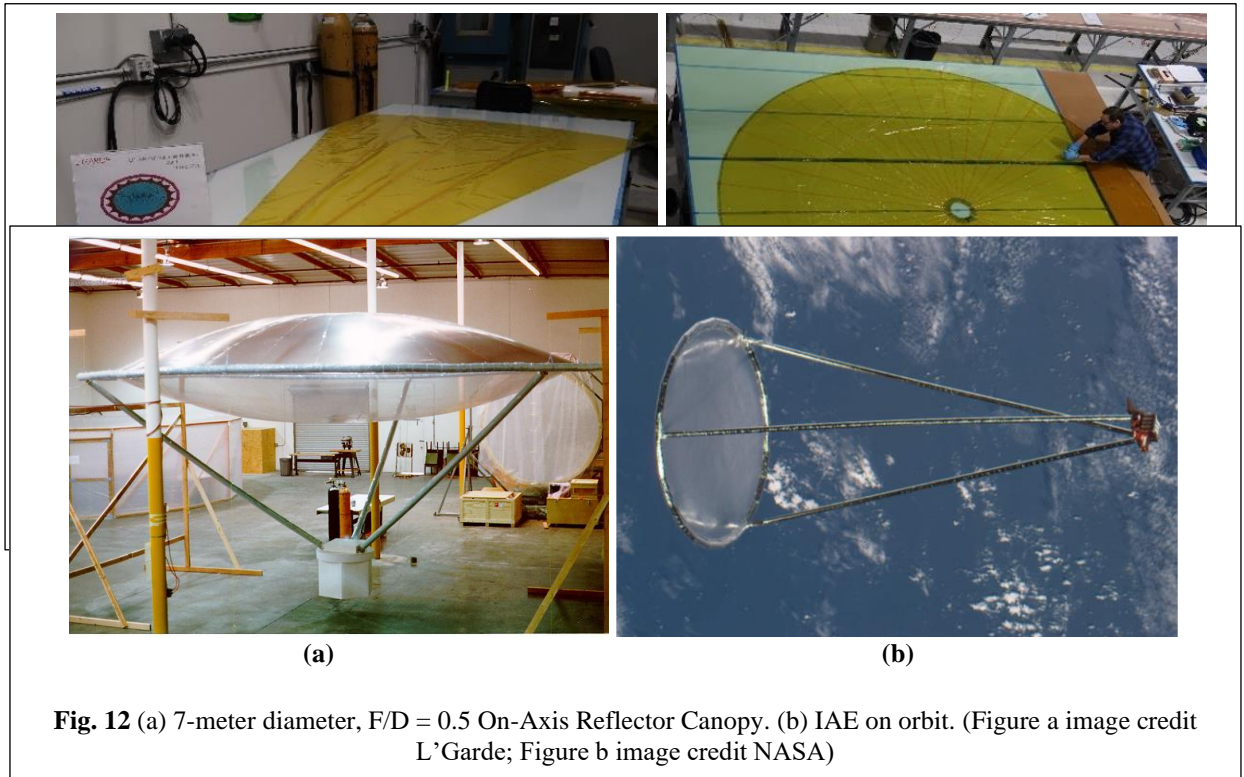

**(a)**                    **(b)**

**Fig. 12** (a) 7-meter diameter, F/D = 0.5 On-Axis Reflector Canopy. (b) IAE on orbit. (Figure a image credit L'Garde; Figure b image credit NASA)

We have designed, fabricated and measured antenna reflector diameters ranging from ½ meter to 1 meter [24], 3 meter [25], 7 meter [26], and 14 meter [17]. The 7-meter reflector is shown in Fig. 12(a) and the 14-meter IAE reflector on-orbit is shown in Fig. 12(b).

## 2.4 Expected Performance

We have essentially eliminated the presence of localized deviations due to the sealing lines by several means. In the design of the gores, we analytically consider the presence of thicker seam tapes at the sealing line areas in our calculation of the gore shape of the initial uninflated configuration. This means that on inflation to the desired pressure, $P_0$, we obtain a surface figure very close to that desired – in this case, a paraboloid. By design, the inflation pressure value is high enough to give membrane strains that remove the packaging wrinkles and creases. Reflectors designed in this manner and made by L'Garde, show no visible distortions at the seams nor at the perimeter. This was the case on the 14m diameter IAE as well as subsequent inflatable paraboloid reflectors. More recently, a 3m and 1m diameter inflatable paraboloid reflectors we designed and fabricated show the same wrinkle free behavior.

SALTUS will use a highly accurate 3D Gore Mandrel Assembly that can thermoform the initial uninflated gore shapes rendering them with curvatures not only along the meridian but along the hoop direction as well. Use of thermoforming results in the M1 uninflated shape being very close to the theoretical starting configuration, but also results in continuous optical surface across sealing lines. Moreover, to make the initial shape even closer to the theoretical curvature, we use a "seamless approach" in conjunction with thermo-formed gores. Seamlessness is achieved by using two sets of identical gores that are bonded in a staggered manner. The thermoforming of M1 gores results in the minimization of membrane thickness variation over the M1 surface.

Optical simulations show that a SALTUS M1 mirror with up to about 7 mm RMS deviation from the best-fit-surface is correctible to the diffraction limit of the SALTUS wavelengths of interest. [14] The inflatable reflectors we have made to date, including the IAE, have surface accuracies better than this. Moreover, with our new fabrication method using the 3D gore mandrel assembly, we expect to achieve surface accuracies 5 to 10 times better than the current state of the art, below 1 mm RMS.

From a past L'Garde project called the "Large Inflatables Structures (LIS)", analysis that showed for a 21K temperature gradient across the surface of a 15m diameter reflector, the surface degradation is on the order of 2.5 mm RMS. This level of M1 surface distortion is correctible to the diffraction limit of the SALTUS wavelengths of interest.[14] Moreover, since the low temperature and low temperature variation over the M1 surface can be known by analysis and experiments, their deleterious effects on the M1 accuracy can be designed out. The low temperature gradients help to minimize the negative effects of the CTE and CTE non-uniformity over the M1 continuum.

We have performed experiments on subscale inflatables to determine the minimum film stress needed to remove wrinkles and creases due to packaging. The results of our experiments show that the film stress must at least be 4.5 MPa (650 psi). SALTUS' design pressure of 5.1Pa results in a 6.89 MPa (1,000 psi) film stress to assure no packaging wrinkles and creases remain. Further, we will consider the creation of wrinkle and creases as we detail the stowing approach, see Fig. 13. [27]

Measurements of the surface accuracy of inflatable paraboloid reflectors using Lidar and photogrammetry, show that the surface accuracies achievable on SALTUS and IAE-like diameters are on the order of 1mm RMS. This was achieved after the reflector membrane has been folded and packaged.

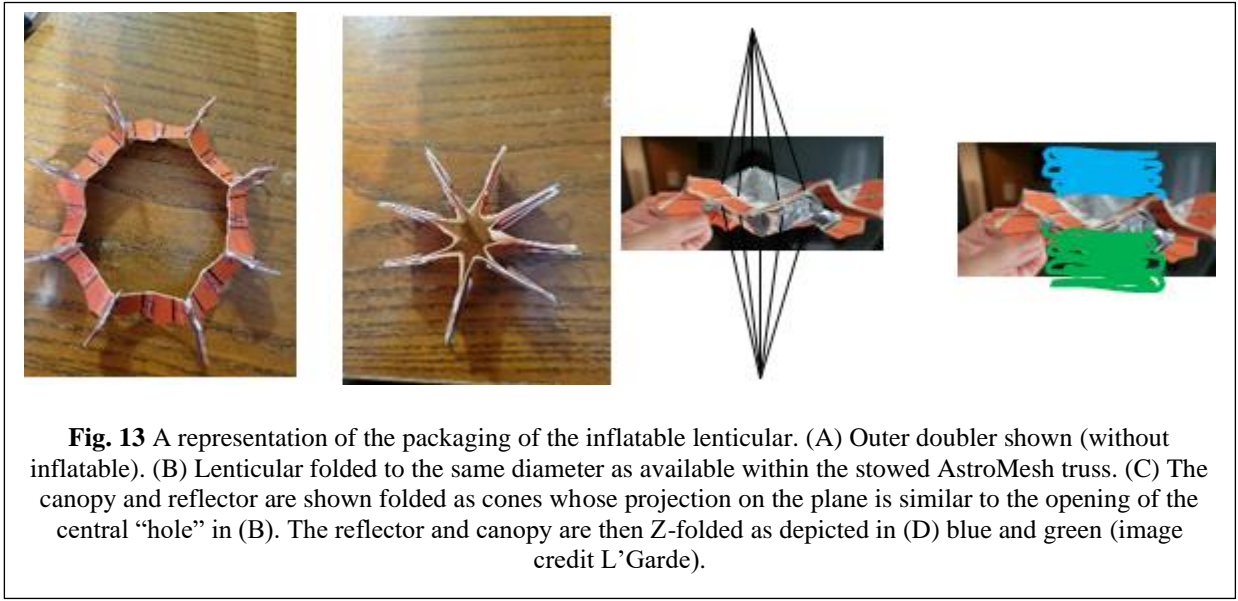

**Fig. 13** A representation of the packaging of the inflatable lenticular. (A) Outer doubler shown (without inflatable). (B) Lenticular folded to the same diameter as available within the stowed AstroMesh truss. (C) The canopy and reflector are shown folded as cones whose projection on the plane is similar to the opening of the central "hole" in (B). The reflector and canopy are then Z-folded as depicted in (D) blue and green (image credit L'Garde).

### 2.5 Stowage and Deployment

A concept for the packaging of M1 is notionally described in Fig. 13. The outer annular ring in Fig. 13 (a) is analogous to the scalloped perimeter doubler for the 3m inflatable shown in Fig. 9 (a). This doubler is folded as shown in Fig. 13 (a) and Fig. 13 (b), while at the same time the canopy and reflector are suspended above and below using fabrication support equipment (FSE) as shown in Fig. 13 (c). After folding to the stowage size, the edges of the perimeter doubler are mated with the inner AstroMesh™ perimeter. Another approach we are investigating is to first attach the inflatable lenticular to the inner AstroMesh™ perimeter and fold both the inflatable and AstroMesh™ in synchrony. The inflation deployment of the SALTUS inflatable is preceded by the full deployment of the AstroMesh™ perimeter truss. Inflation to full pressure is initiated only after the perimeter truss is fully deployed and able to take load.

### 2.6 Contamination

Far-IR systems can be quite sensitive to contamination, especially if contamination increases the emissivity of a surface. In the case of M1, we must be concerned with contamination from the manufacturing process and ground handling as well as during flight. The main concern in flight will be particles and molecular contamination, hydrocarbons and water from the inflatant tanks. Fig. 14 shows the absorption cross section, $Q_{abs}$, from a Mie theory calculation for particles of salt (NaCl), carbon and aluminum at a wavelength of 100 μm. The three chose species represent spit and sweat from human activity, carbon is a surrogate for general soot and dust and the aluminum represents metallic debris. The emissivity of a particle is proportional to $Q_{abs}$ and the fractional area coverage (FAC). As can be seen from the plot, the larger particles have larger absorption cross sections and therefore are the greatest potential contamination hazard in terms of emittance. We will manufacture M1 in a clean environment and limit the particles that are on the membranes as it is assembled. Further, techniques were developed to clean membranes of the size of M1, namely the James Webb Space Telescope (JWST) sunshield layers, which had a measured coverage of 0.002-0.007 percent area coverage (PAC=FAC*100). [28] On this basis, we firmly believe we can manufacture clean membranes.

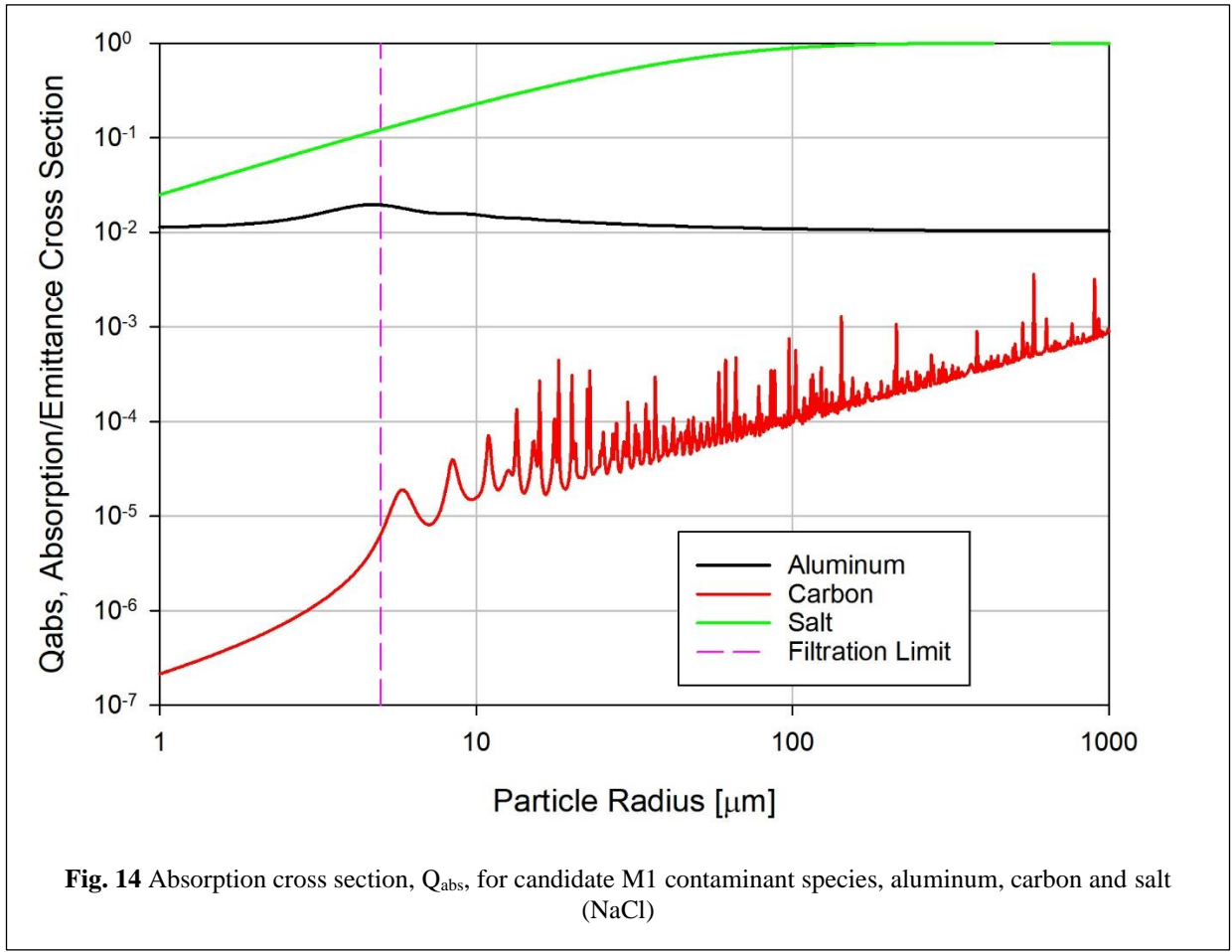

**Fig. 14** Absorption cross section, $Q_{abs}$, for candidate M1 contaminant species, aluminum, carbon and salt (NaCl)

It is well known in contamination control engineering that surface cleaning preferentially removes the large particles, which is the desire for this application.

In order to limit the particles inside M1, the ICS will use ultra-high purity He inflatant and will have a filter to limit particle from being pushed into the volume of M1 over the life of the mission. This filter can be seen in the high pressure section of the ICS block diagram (Fig. 15), limiting particle size to about 5 µm, indicated by the vertical red line in Fig. 14 and shows that the $Q_{abs}$. will be limited to small values, via limitation of particle sizes.

Table 1 provides a summary of the impacts of contamination from all sources and the planned mitigation. In short, the contamination requirements and mitigations will ensure the performance of M1 is not compromised.

**Table 1** Contamination impacts and mitigations

| Contamination Type (Requirement) | Impact on SALTUS | Mitigation |
|---|---|---|
| **Particulate (0.5 PAC/surface)** | - Emissivity and scattering increase., transmission loss<br><br>- Dominated by larger particles and those whose sizes are the same as SALTUS wavelengths [29] | - M1 is not cleanable after assembly so strict protocols for manufacture, storage, and monitoring<br>- NB: Cleaning preferentially removes large particles so impact very small on latter optics in the corrector module[14] |
| **Water Ice (20 Å/surface)** | - Transmission loss due to absorption loss mainly from 30-100 μm. 20 Å/surface is 1% peak loss (~45 μm) [30]<br>- Increased emittance in low emittance (coated) surfaces [31]. This increase in emittance will be an increase in radiance for elements with limited sky view [32]. | - Limit ice accumulation through design. Design sources, like composite structure to cool before optics or add heaters to optics. (Legacy JWST) Plan proper venting and contamination analysis to predict icing. |
| **NVR (30 Å/surface)** | - Hydrocarbon layers from long term exposure expected to be less than 10 Å [31, 32]. Such a small layer at a reflecting surface has negligible optical impact [28]. | - Flight hardware will be kept in a clean filtered environment. |

## 2.7 ICS Architecture

The ICS is a closed loop control system that regulates the flow of inflatant (helium) from high-pressure storage tanks on the spacecraft through gas lines on the boom to M1. It maintains the pressure within M1 to meet performance requirements over all attitudes and through the mission during science operations. The ICS maintains reflector pressure during science observations within 0.001 of the design pressure, $P_0$, 5.1Pa, 5.1 mPa.

The pressure in M1 is sensed through multiple strain gauges made from piezo-electric film bonded to M1 connected in a 4-wire arrangement and located outside the optical clear aperture. The strain gauge produces a voltage proportional to film strain, which is proportional to pressure and forms the basis of the error signal for the control loop. Our current baseline is to operate this loop as a bang-bang controller, other control algorithms will be studied in Phase A. Based on the error signal, gas will either flow into M1 or be released from it.

The ICS architecture is presented in Fig. 15. The reservoir of 4 ARDE 4992 tanks, protected by Kevlar overwrap holding to 50kg helium each at 4000 psia. [2] The latch valves, LV1 and LV2 are opened, a regulator passes filtered helium to the low-pressure stage. This lower pressure stage is plumbed to a series of proportional flow control valves (PFCV) independently controlled using the 3-channel spacecraft torque control resources, 12-bit regulated current source. Each PFCV is followed by a flow control orifice, which are sized to produce three channels with a dynamic range of approximately 6.46, with a 10% flow overlap, giving the ICS an overall dynamic range of approximately 218. The upper mass flow rate is set by the maximum gas loss rate at EOL and for the most demanding environment, ~0.9 mmol/sec. This arrangement gives a flow resolution of 6 nmol/sec.

The ICS has four operating modes; (1) inflation, (2) discontinuous inflation, (3) continuous inflation and (4) release. Mode 1, the inflation mode is used after deployment of the boom and truss, where the inflation rate is slowly adjusted to the maximum rate and completes in about 40 hours depending on the flight ambient temperature. The boom and truss are the first to deploy and the sunshield layers last. [2] Mode 2, the discontinuous inflation mode is required when the M1 loss rate is below the minimum continuous flow rate of approximately 3.7 μmol/sec. For the discontinuous inflation mode, a duplicate to the lowest flow channel is employed (indicated in Fig. 15 by the cyan box) and a valve is opened for a fixed time (3 sec) and then closed for an interval which decreases as the M1 loss rate increases with mission elapsed time. An additional valve is needed as the number of cycles for Mode 2 is about 70% of valve life as specified by the vendor. At about 9 to 12 weeks the expected MM flux will increase the loss rate to above the minimum continuous flow capability of the ICS. After the mean gas loss rate is above the ICS minimum continuous flow capability, the ICS will transition to Mode 3, continuous flow, and will remain in this mode for the rest of the mission. The ICS has sufficient resolution that the error between the continuous loss rate of M1 and the discrete inflow rates have a mean error of a few nmol/sec, allowing the ICS control loop to maintain pressure to within requirements.  The final mode, Mode 4, release, is needed when SALTUS changes attitudes, that decreases the sun angle and heats M1. The  temperature change is proportional to the change in the 4th root of the change in cosine of the Sun angle, it will cause an over inflation, much greater than pressure tolerances allow. Excess inflatant is vented by opening on of the (redundant) drain valves (C or D), see Fig. 15.  The vent diameter is sized such that the maximum release time is 20 minutes and will occur during slews. In a later more detailed design phase of the program, the vent diameter will be optimized to match the final slew rate of SALTUS so as not to impact observational efficiency.[2] The venting resulting from cold to hot point is discussed in more detail in 3.4.3.

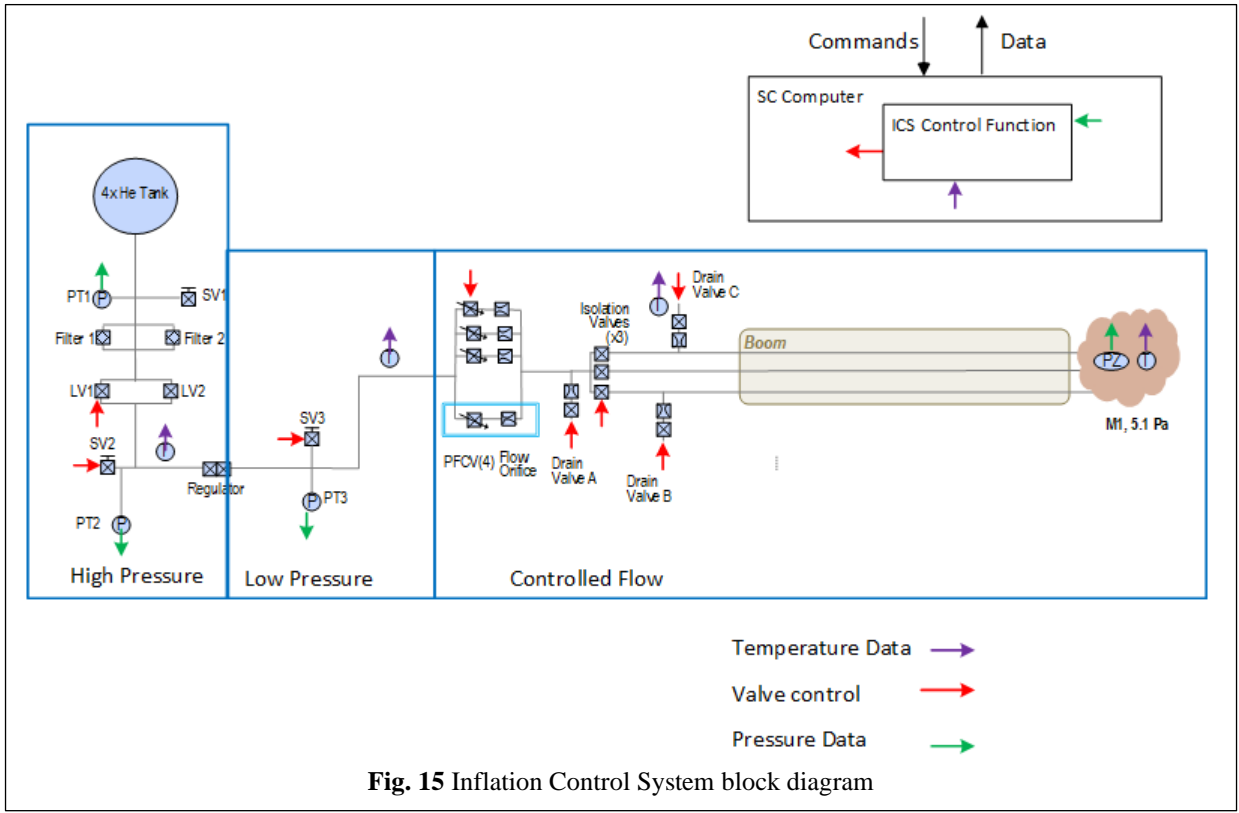

**Fig. 15** Inflation Control System block diagram

## 3   M1 in the Operating Environment

SALTUS will operate in a halo orbit near Sun-Earth L2. [2] The SALTUS environment is for our purposes interplanetary space, the main environmental challenges come from, solar radiation, wind, and micrometeoroids (MM). This section discusses how we have modeled M1 and concludes with a discussion of the impact of solar light pressure, wind, and MM on the shape of M1.

*3.1 Thermal Modeling of M1*

M1 is essentially isolated in the sense of thermal conductivity. M1 is very low mass and therefore, in effect, has no thermal transient. The temperature of M1 is dependent on the temperature of the top layer of the sunshield and can be determined almost exactly with the Planck function and a little geometry. We seek a simple form to understand the drivers on the temperature of M1. As a first order approximation we consider M1 enclosed by a sphere, this sphere has two sections; the top hemisphere is open and exposed to cold (4K) space and the bottom hemisphere that sees the top layer (L2) of the sunshield. Let F represent the fraction of the full solid angle subtended from M1 by the sunshield, in this representation 1-F is then the fraction of the solid angle of cold (4K) space that M1 sees.

We seek to know what the M1 mirror temperature, $T_{M1}$, is a as a function of the temperature of the sunshield Layer 2 (top or coldest layer), $T_{L2}$, shield and mirror emissivities, $\varepsilon_{L2}$ and $\varepsilon_{M1}$ and F. Since M1 is conductively isolated and the system is in equilibrium we know that

$$heat\ in = heat\ out\ .\tag{1}$$

The equilibrium condition (1) can be written as

$$A_{M1}\alpha_{M1}\left(F\varepsilon_{L2}\sigma T_{L2}^4\right) = A_{M1}\varepsilon_{M1}\sigma T_{M1}^4\left(1-F\right)\tag{2}$$

where $A_{M1}$ is the area of M1, $\alpha_{M1}$ is the absorptance of M1 and $\varepsilon_{L2}$ and $\varepsilon_{M1}$ are the emissivities of the sunshield Layer 2 and the M1 respectively. Cancelling like terms in (2) gives

$$\alpha_{M1}\left(F\varepsilon_{L2}\sigma T_{L2}^4\right) = \varepsilon_{M1}\sigma T_{M1}^4\left(1-F\right)\ .\tag{3}$$

Solving (3) for $T_{M1}$ gives

$$T_{M1} = T_s \sqrt[4]{\left(\frac{\alpha_{M1}}{\varepsilon_{M1}}\right)\varepsilon_{L2}\left(\frac{F}{1-F}\right)}. \qquad (4)$$

For this simple analysis we will use the so-called gray approximation, namely $\alpha_{M1} = \varepsilon_{M1}$. Under this approximation, (4) becomes

$$T_{M1} = T_s \sqrt[4]{\varepsilon_{L2}\left(\frac{F}{1-F}\right)}. \qquad (5)$$

In Fig. 16, (5) is solved for $\varepsilon_{L2}$ in the (Ts, F) plane for various values of $\varepsilon_{L2}$ when $T_{M1}$ is 40 K. For the current baseline of SALTUS, F~0.1 and $T_{L2}$ has a maximum ~80K, this design point is to the left of all the contours meaning that the emissivity of 0.7, that of a black membrane would result in an acceptable M1 temperature.

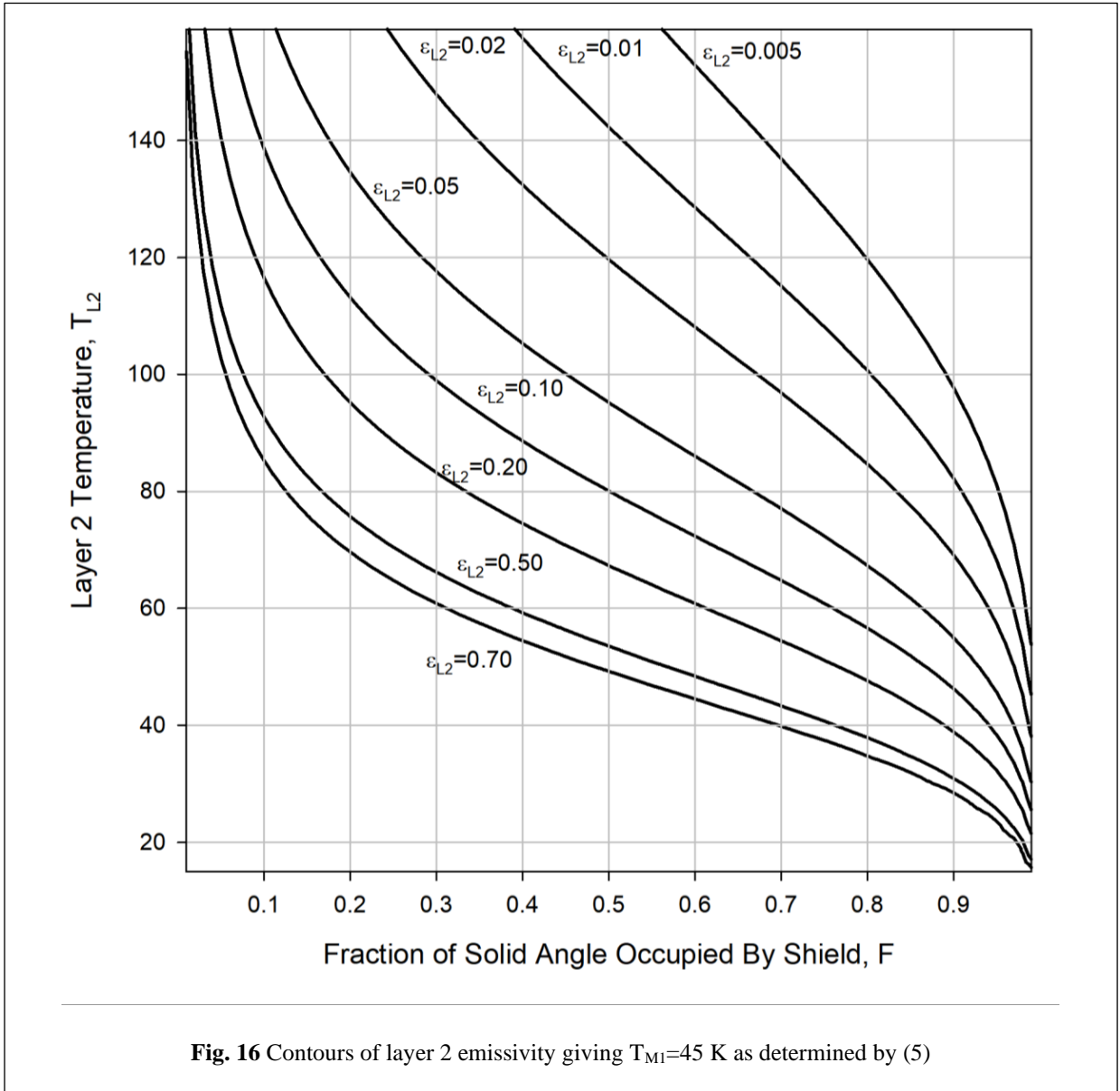

**Fig. 16** Contours of layer 2 emissivity giving $T_{M1}=45$ K as determined by (5)

The results in Fig. 16 illustrate the wide range of possible successful architectures for SALTUS. The flexibility in the fundamental thermal architecture gives us confidence that as the concept matures, we will continue to be able meet and perhaps exceed temperature requirements and deliver even greater sensitivity for the SALTUS mission.

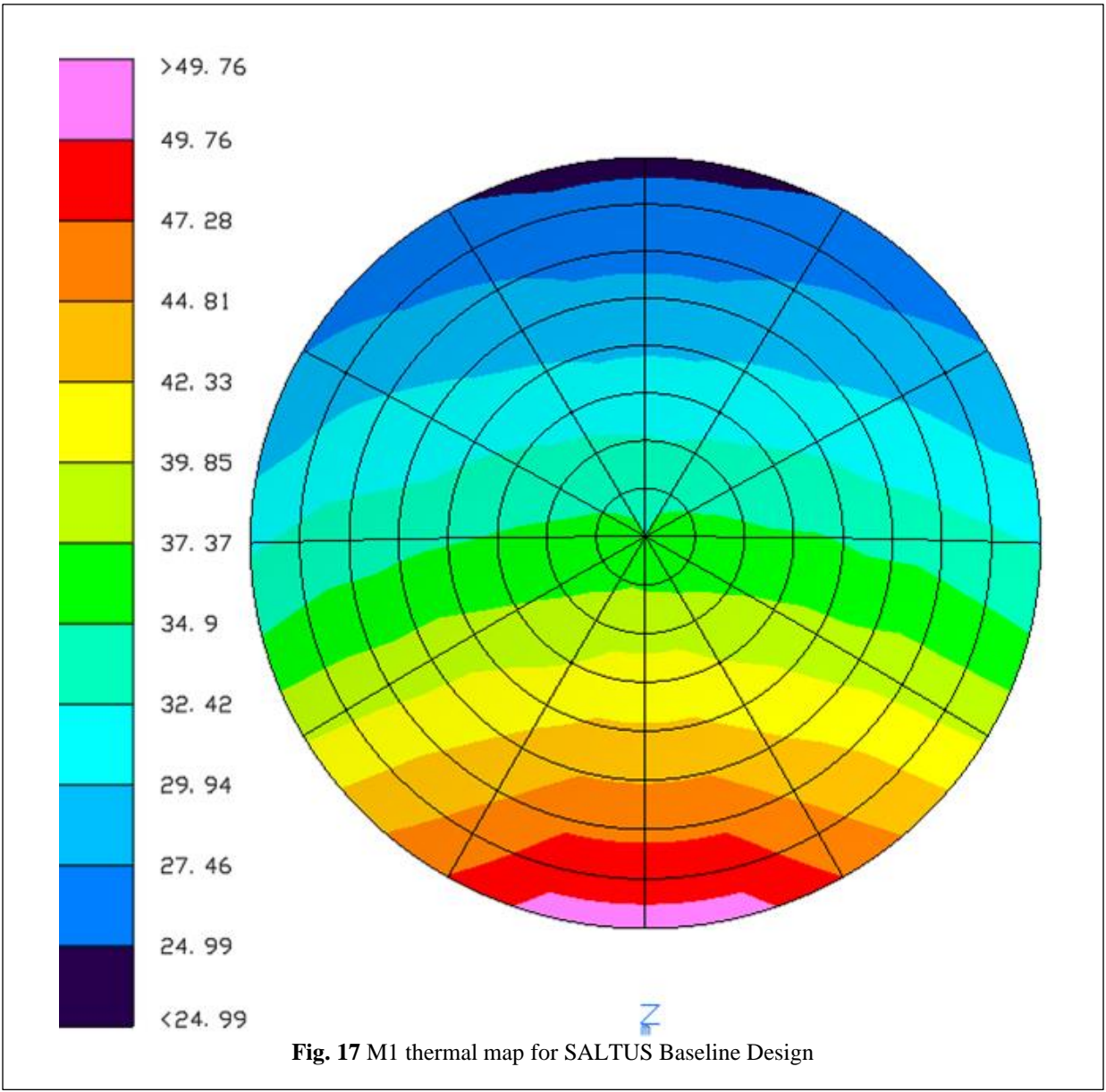

**Fig. 17** M1 thermal map for SALTUS Baseline Design

We subsequently built a more traditional thermal model which produced the M1 temperature shown in Fig. 17. In addition to the proposed baseline architecture, four other cases were run. For each of these cases the maximum, minimum and mean M1 temperature were determined and plotted against the maximum temperature of sunshield Layer 2. These results are plotted in Fig. 18, and these data fall nicely on straight lines verifying the results given in (5).

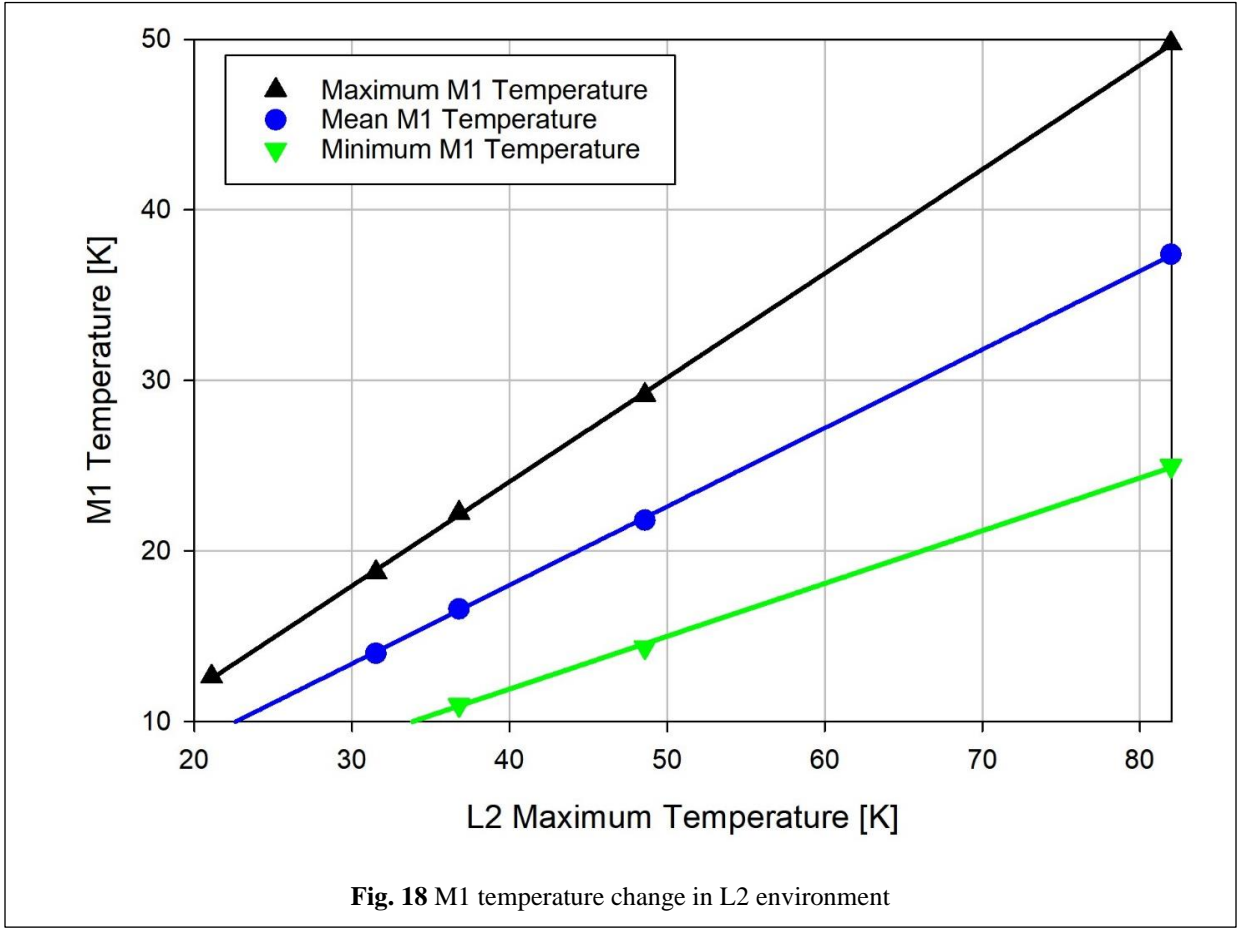

**Fig. 18** M1 temperature change in L2 environment

*3.2 Interaction With Solar Emission*

Photons and plasma from the Sun could pose a problem to M1 in two ways: the first is by applying external pressure on the inflatable. Second, accumulation of charge on M1's surface could result in electrostatic discharge that damages the observatory.

Pressure on M1 can come from two sources, pressure from the solar wind (the outflow of protons and electrons from the sun) and the pressure exerted by the Sun's light.

At the L2 Lagrange point, SALTUSs operational milieu, M1 is embedded in the supersonic solar wind, which is composed of roughly equal parts protons and electrons traveling radially outward from the Sun. The dynamic pressure of the solar wind on an obstacle can be calculated as $p = m_p \rho V^2$, where $m_p$ is the proton mass, $\rho$ is the proton number density, and V is the solar wind bulk velocity. We use data from the WIND spacecraft, a solar wind monitor located at L1, to estimate this effect on M1. We focus on data from solar maximum during the year 2001, when extreme solar wind disturbances are most likely to occur. We find that the maximum solar wind dynamic pressure recorded by the spacecraft was ~500 nPa, which corresponds to pressure many orders of magnitude below the pressure tolerance of M1, $\delta P = 5 \times 10^{-3}$ Pa. Therefore, we conclude solar wind dynamic pressure effects can be neglected.

Pressure induced by solar radiation is I/c where I is the solar intensity of approximately 1,370 W/m² and c is the speed of light. This gives the peak pressure due to solar radiation as 4.56 µN/m² or 4.56 µPa, a value even smaller than the pressure due to the solar wind, so also negligible.

Both the effect of solar wind and light pressure were evaluated without considering the effect of the sunshield. The presence of the sunshield can only improve matters and we need no requirements on the performance of the sunshield as a solar windbreaker as the even full wind impact is negligible.

Since M1 will be exposed to the solar wind and if proper design mitigations are not taken, charge could accumulate. Furthermore, the sunshield's shadow precludes sunlight dissipating charge build up on M1.

The accumulation of charge could deform M1 through Lorentz forces as M1 moves through the sun's electromagnetic field or damage M1 through arcing. It may be possible to show these effects are small, but such analysis is challenging and uncertain as the environment is not well known [17] and therefore a resource burden on the program. Consequently, we will take active design measures to avoid the build-up of charge by employing dissipative coatings and connecting M1 to the system ground. This global ground also avoids the risk of static charge build up and the risk that such static cling will impede deployment. The current baseline calls for the incorporation of grounding features along seam lines. A conceptual sketch is shown in Fig. 19.

The grounding of a large membrane system has previously been implemented, with even more surface area than M1, namely the JWST sunshield. [35] SALTUS will use similar design practices to JWST and apply any relevant lessons learned.

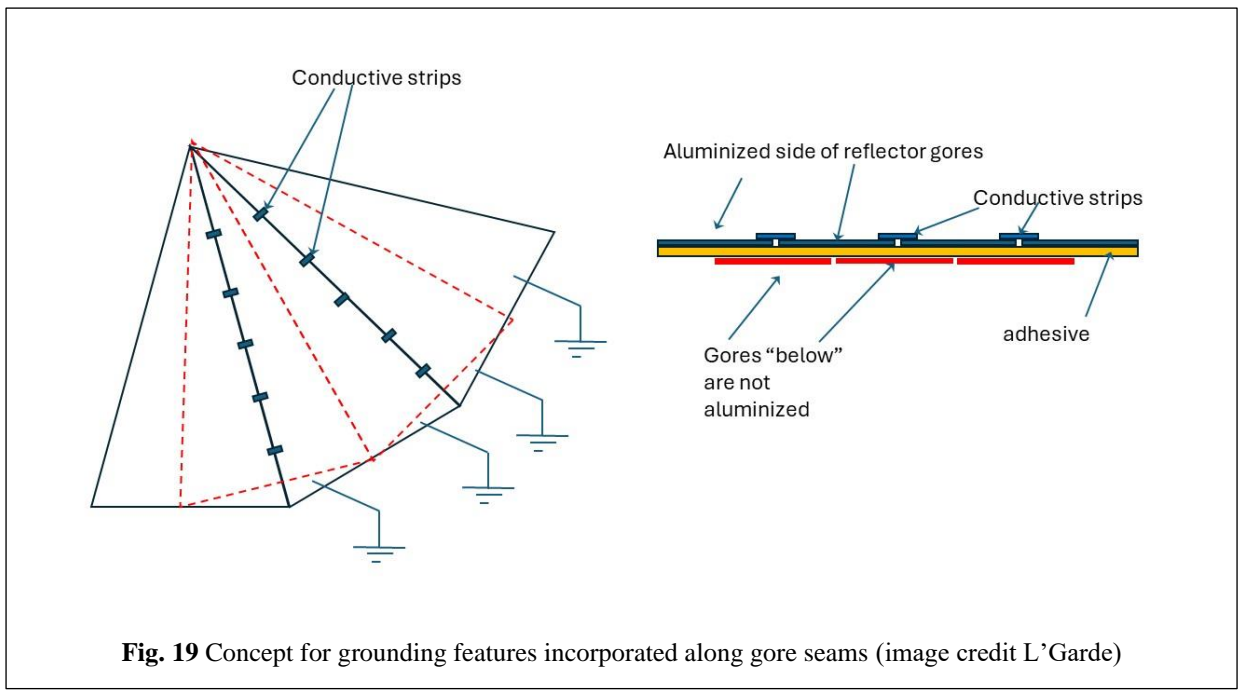

**Fig. 19** Concept for grounding features incorporated along gore seams (image credit L'Garde)

*3.3 Micrometeoroids*

MM affect M1 through two imprecisely known factors; the first is the environment or MM distribution at L2 and the second is the damage caused by interaction of a MM of a given size and energy. Our work on understanding each of these factors will be discussed in the initial section, followed by discussion of the impact of potential gas leaks and critical time scales.

### 3.3.1 Micrometeoroid Environment

MM pose a threat to our inflatable primary reflector M1 as the impacting particles will create punctures in the surface, resulting in gas loss. Since the gas loss rate will partially determine the mission lifetime, it is crucial to properly estimate the impact rate at its L2 Lagrange point location. One tool often applied for mission planning is SPENVIS (https://www.spenvis.oma.be/), which can provide the integral MM flux above a threshold mass. This model adopts an analytical description for the interplanetary MM flux near 1 AU from Grün et al. (1985) which is valid for particle masses $10^{-18} < m < 1$ g. [36]

It is important to note that this is a "sky-averaged" description for the integral flux. In their paper [36], Grün et al. acknowledge that the sky-distribution is anisotropic and applied corrections to data taken from reference spacecraft which pointed in directions of enhanced MM flux (i.e., Earth's Apex) to arrive at this result. In this section, we compare the SPENVIS model to recent models and in-situ data for expected fluxes at the L2 location. In addition, we also consider the pointing direction of M1 in the context of current assumptions and derive a correction to the SPENVIS model for the SALTUS design.

Recent MM models rely on orbital elements of the parent body source to predict fluxes at specified solar system locations. Thorpe et al. [37] consider contributions from Jupiter Family Comets (JFC), Halley-type Comets (HTC), Oort Cloud Comets (OCC), and Asteroids (AST). They develop an independent description for the sky-averaged integral flux at the L1 location listed in their Table 2. These equations are expressed as a function of the threshold particle momentum, p.

In addition to the integral flux, Thorpe et al. [37] also show the predicted angular flux density from each population source for a momentum threshold of ≥ 1 μNs (reproduced here as Fig. 20). These sky maps demonstrate that the various populations produce an anisotropic sky distribution which has its largest concentrations in flux in the sunward and anti-sunward directions, with additional notable contributions in the north/south pole directions and near the prograde direction. The sunward and anti-sunward fluxes are dominantly sourced from Jupiter Family Comets (JFCs) which are expected to account for ~90% of the MM particles at L1, and similarly L2. [37] Since the surface normal of M1 will primarily align with the sunward/anti-sunward direction, we

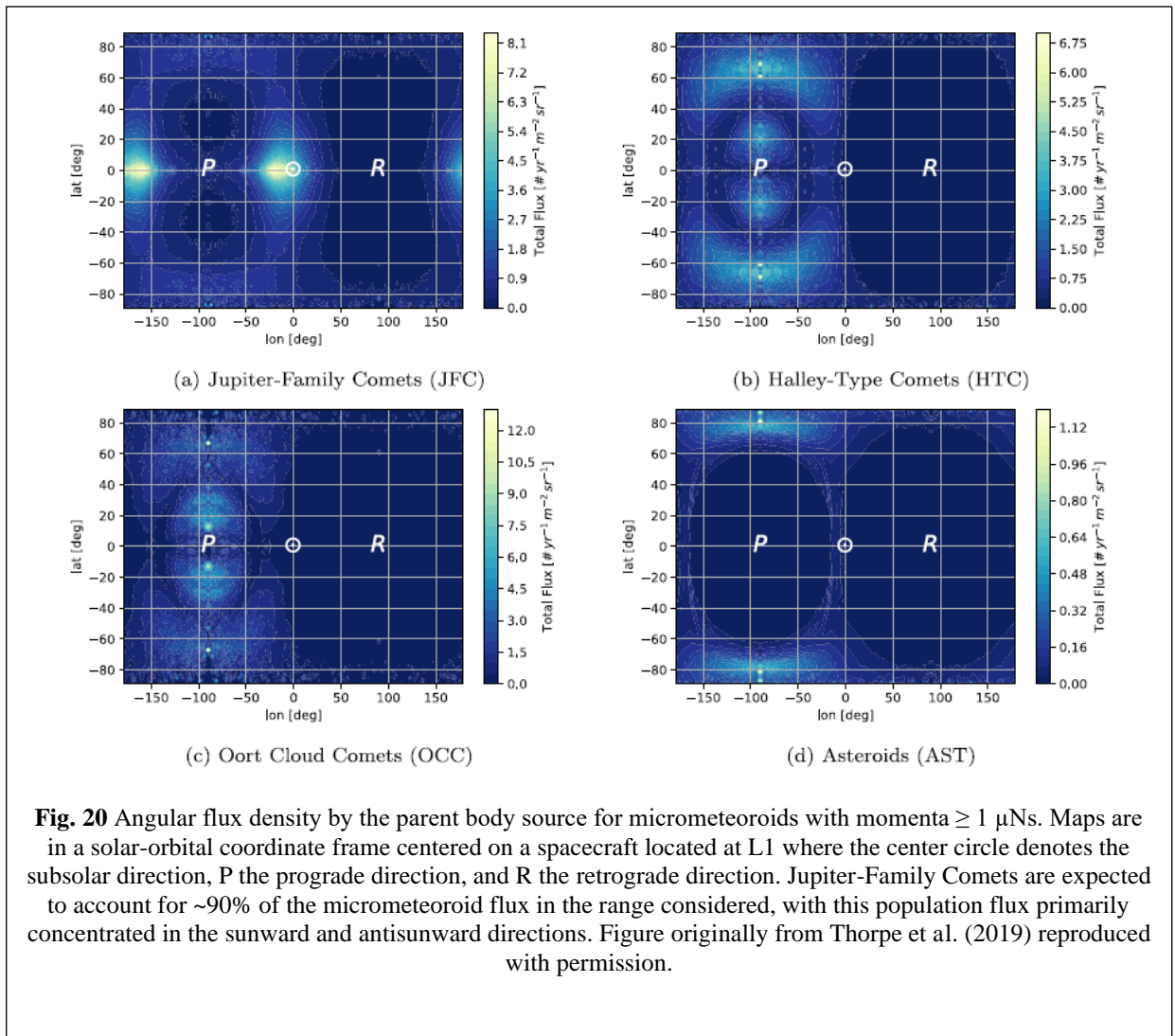

(a) Jupiter-Family Comets (JFC)

(b) Halley-Type Comets (HTC)

(c) Oort Cloud Comets (OCC)

(d) Asteroids (AST)

**Fig. 20** Angular flux density by the parent body source for micrometeoroids with momenta ≥ 1 μNs. Maps are in a solar-orbital coordinate frame centered on a spacecraft located at L1 where the center circle denotes the subsolar direction, P the prograde direction, and R the retrograde direction. Jupiter-Family Comets are expected to account for ~90% of the micrometeoroid flux in the range considered, with this population flux primarily concentrated in the sunward and antisunward directions. Figure originally from Thorpe et al. (2019) reproduced with permission.

anticipate it will be exposed to the region in the sky which is expected to have the highest flux of incoming MM.

In-situ experiments have also been performed to track the rate of MM impacts on spacecraft. LISA Pathfinder was used as an omnidirectional detector of such impacts and found the direction and frequency of sensed impacts was consistent with the sky distribution shown in Fig. 20. [37] Similarly, a study was conducted for the Genesis spacecraft which was at the L1 location for 2.33 yrs. This was based on an aluminum plate that pointed exclusively in the sunward direction. [38] This study identified 32 confirmed impacts on a usable surface area of 220 cm², with an additional

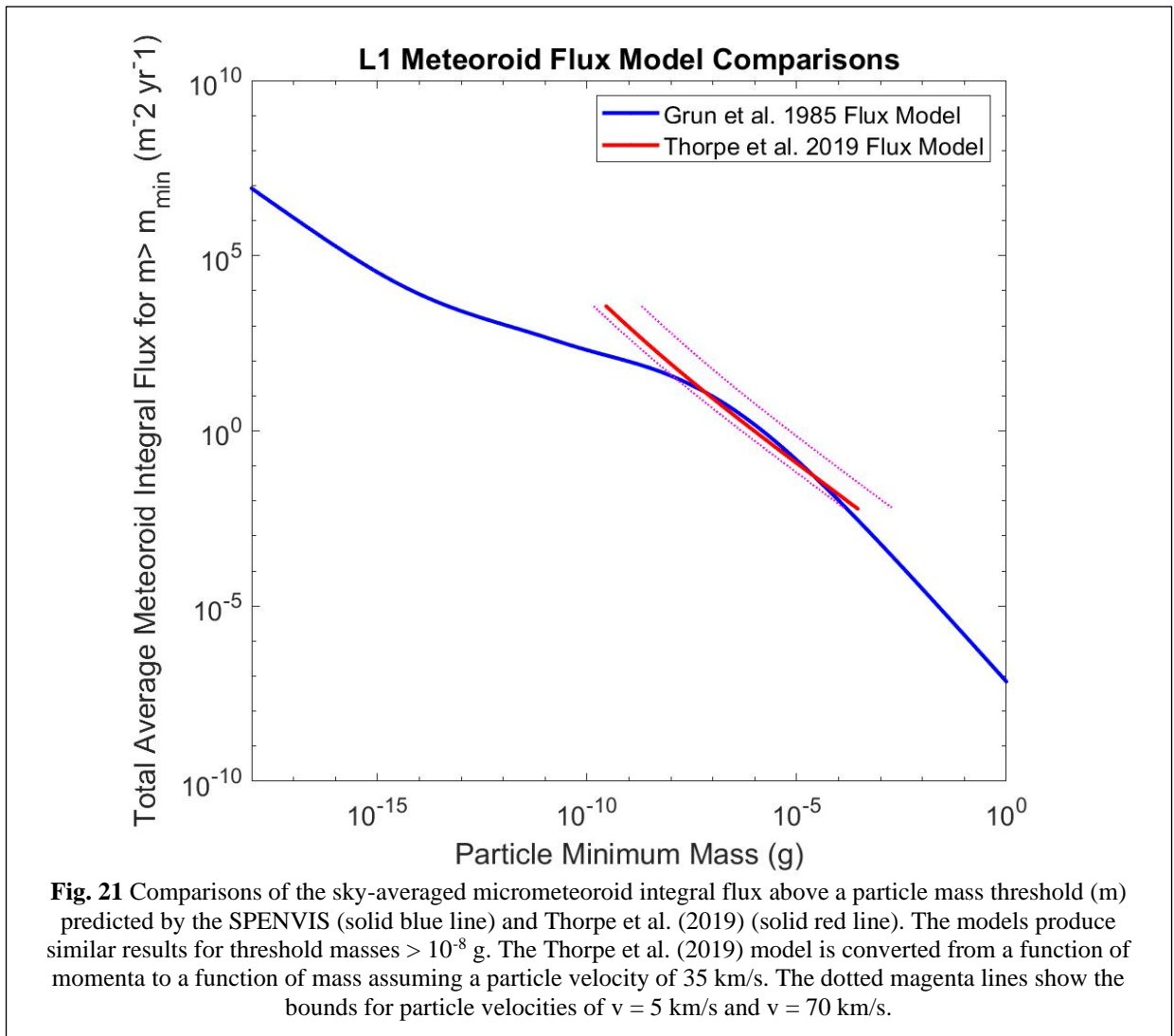

**Fig. 21** Comparisons of the sky-averaged micrometeoroid integral flux above a particle mass threshold (m) predicted by the SPENVIS (solid blue line) and Thorpe et al. (2019) (solid red line). The models produce similar results for threshold masses > $10^{-8}$ g. The Thorpe et al. (2019) model is converted from a function of momenta to a function of mass assuming a particle velocity of 35 km/s. The dotted magenta lines show the bounds for particle velocities of v = 5 km/s and v = 70 km/s.

12 features classified as possible impacts. [38] Given that M1 will be exposed to the same region of the sky, we compare the Genesis results to predictions by the SPENVIS model in the next section.

To directly compare the Thorpe et al. [37] model to SPENVIS, we need to convert the integral flux from a function of momenta to one of mass. To do this, we use the momentum equation p=mv, where p is the particle's momentum, m the particle mass, and v the particle velocity. We assume a meteoroid speed of 35 km/s for all particle masses. Based on the momenta range for which these equations were derived, this would correspond to a threshold particle mass range $\sim 3 \times 10^{-10} < m < 3 \times 10^{-4}$ g.

Fig. 21 shows a direct comparison of the sky-averaged integral flux of MM above a particle mass threshold (m) based on the SPENVIS model (blue line) and the Thorpe et al. (2019) model (red line). The figure shows that these models produce similar results for sky-averaged fluxes near L1 for masses $> 10^{-8}$ g.

We now consider deviations to the predicted M1 MM flux that may occur once the pointing direction is taken into account. The JFC MM sky distribution an angular flux density at the subsolar and antisolar points of ~8 impactors/(m²*yr*sr) for impactors with masses > ~2.8 x 10-8 g. The sky-averaged integral fluxes corresponding to this mass threshold predict 21.4 and 28.4 impactors/(m²*yr) from the SPENVIS and Thorpe models respectively. We can divide these values by the total field of view of the sky, $4\pi$ sr, to arrive at a sky-averaged angular flux density for these models. This results in 1.71 and 2.26 impactors/(m²*yr*sr) predicted from SPENVIS and Thorpe models respectively. Hence, if flux enhancements in the pointing direction are not taken into account, the models may be, depending on the reference model used.

We can also independently calculate a correction factor for the sunward direction by comparing predictions to the observed Genesis micrometeoroid impact counts. The detection method was sensitive to impacting particles with a size > 1 μm.[38] Approximating the particle as a sphere with radius 1 μm and a density of 2.5 g/cm³ [36] this would correspond to a particle mass threshold of m > 1.05 x $10^{-11}$ g that the study could detect. This threshold is below the mass range valid for the Thorpe model, and thus will only be compared to the SPENVIS prediction here.

The aluminum plate onboard Genesis was directed toward the sun for the duration of the mission, with the plate shielded from 25% of the sky by other spacecraft structures [38]. This would give a field of view (FOV) for the plate of 1.5π sr. The Sunward-directed Angular Flux Density can be calculated by:

$$Genesis\ Sunward\ Angular\ Flux\ Density = \frac{32\ impactors}{0.022\ m^2 * 2.33\ yrs * 1.5\pi\ sr}$$

$$= 132.5\ \frac{impactors}{m^2 * yrs * sr}$$

For a mass threshold m = 1 x $10^{-11}$ g and dividing the sky-averaged integral flux by 4π sr, SPENVIS predicts an average angular flux density of 37.5 impactors/(m²*yr*sr), which is below the observed Genesis flux by a factor of ~3.5. Thus, we find the in-situ data is in agreement with the previous section that a correction factor in the range of 3.5 to 5 needs to be applied to the SPENVIS model to properly estimate fluxes on M1.

### 3.3.2 Interaction of M1 with Micrometeoroids

Two things can happen when MM impacts an inflatable. The impactor can puncture M1 and cause loss and or it can fail to penetrate it and cause a mechanical disturbance.

The flux enhancement coming from the solar/antisolar directions (Section 5.1) is written relative to the standard Grün et al. (1985) model as

$$f(\theta) = f_0 + 4f_0 \cos \theta \qquad (6)$$

where $f_0$ represents the omnidirectional (background) flux. We can see from (6) that when $\theta=0$, the surface normal points in the solar direction and consequently the flux is enhanced by a factor

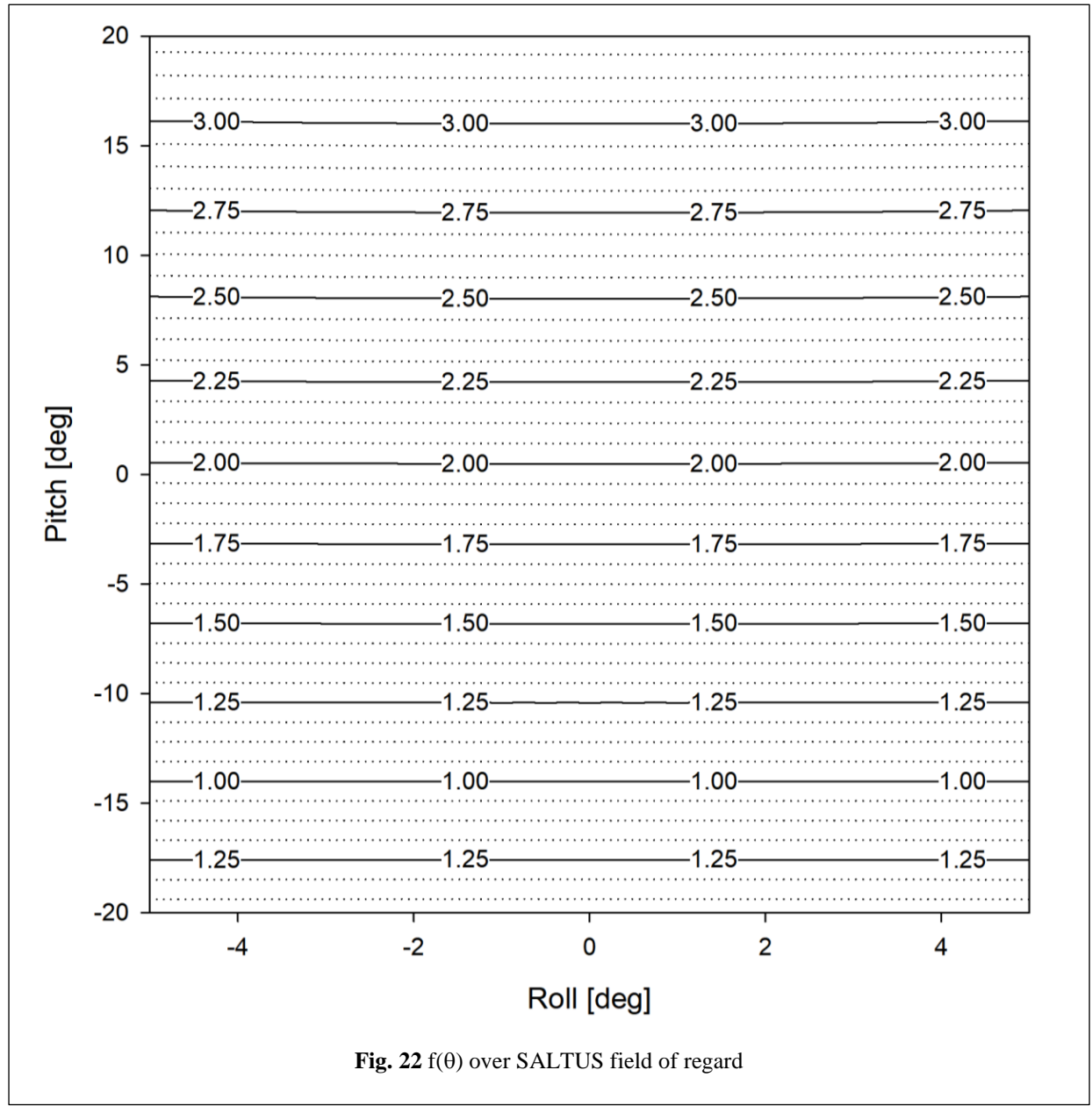

**Fig. 22** f($\theta$) over SALTUS field of regard of 5. However, when the surface normal points perpendicular to the sunward direction, $\theta = 90$, then the surface element is hit by the average background flux. To understand the flux variation of all combinations in between this range, we varied the solar pitch and solar yaw angles each from 0 - 90 degrees. Fig. 22 shows the flux enhancement factor M1's entire FOR for a given pitch and yaw pair. We note here that we have done this calculation assuming the peak flux will be located at the subsolar and antisolar points. However, the incoming direction of the MM as seen from the body frame will be slightly off from the sun vector direction due to motion of the spacecraft. This effect is similar to rain falling directly down on a car, but when viewed from the car frame, appears to fall at an angle when the car is in motion. Future analysis will need to take into account the spacecraft velocity to more accurately calculate the flux enhancement at a given pitch and yaw.

Equipped with a description of the M1 MM environment and the enhancement factors, we can calculate the return period of particles impinging the surface. The number of MM impacts on a telescope can be estimated by

$$C = \varphi\left(m\right)\frac{\pi D_T^2}{2} \qquad\qquad (7)$$

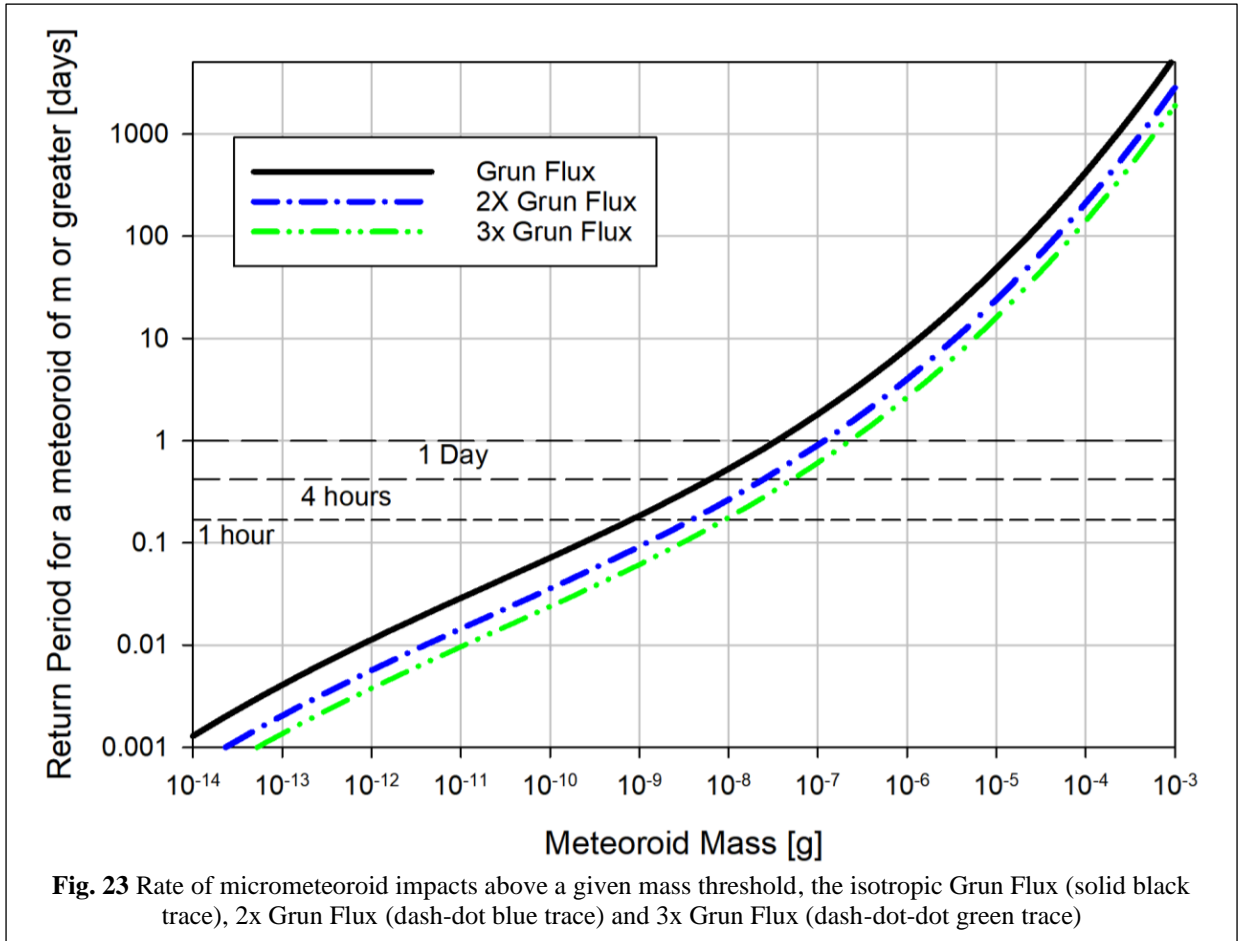

**Fig. 23** Rate of micrometeoroid impacts above a given mass threshold, the isotropic Grun Flux (solid black trace), 2x Grun Flux (dash-dot blue trace) and 3x Grun Flux (dash-dot-dot green trace)

where φ(m) is the micrometeoroid flux and $D_T$ is the telescope diameter. For SALTUS, the diameter of the primary reflector is 15 m and the MM flux, $\varphi\left(m\right)$ is represented by the Grun et al. (1985) model correction factor for non-isotropy of the flux, (7) becomes

$$C = \varphi(m)f(\theta)\frac{\pi}{2}D_T^2. \qquad\qquad (8)$$

We can take the inverse of (8) to analyze the return period for impacts above a mass threshold size. The results of this calculation are shown in Fig. 23. Particles with $m < 10^{-8}$ g impacting may be problematic, depending on their effect on the telescope as they impact about once per hour.

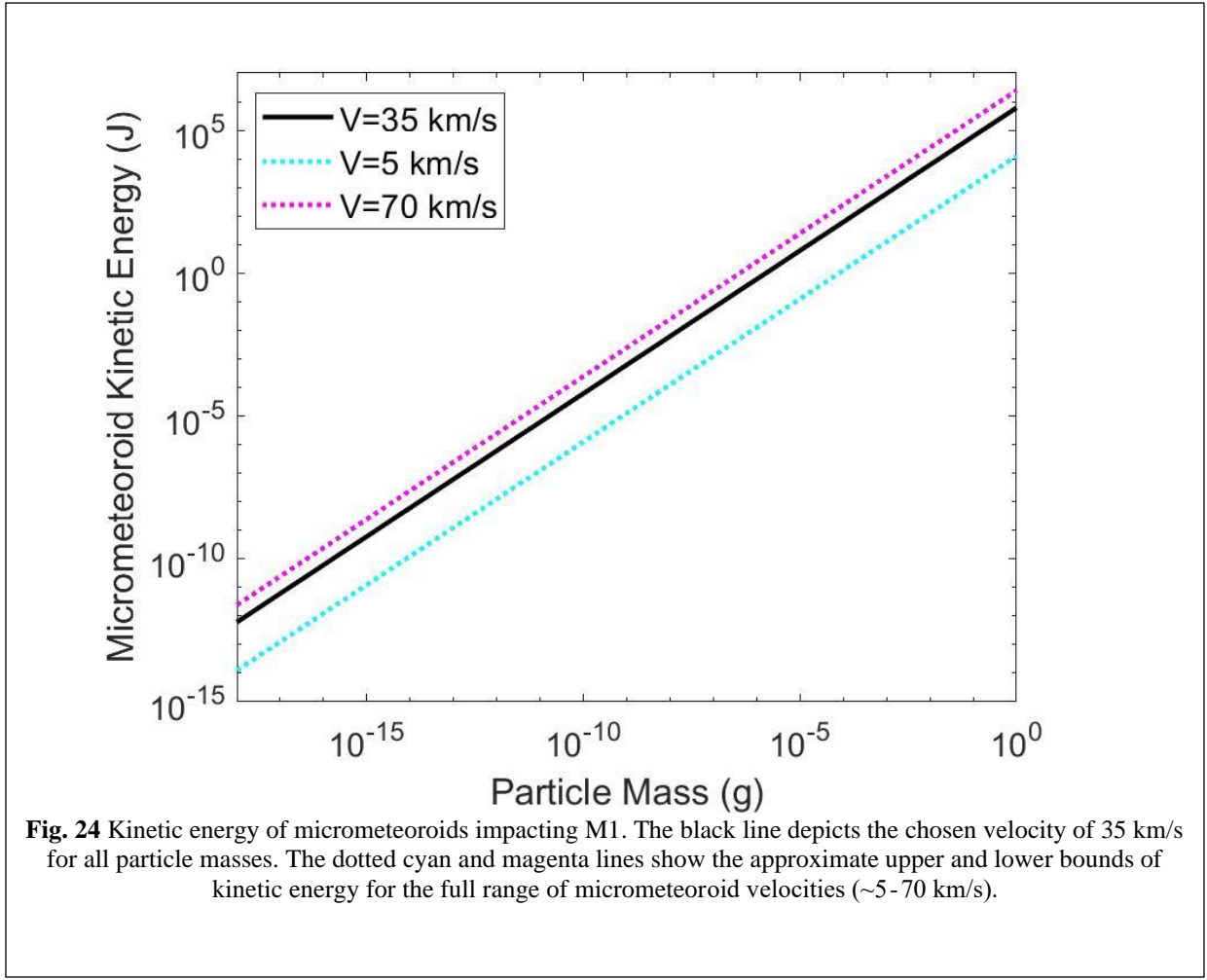

**Fig. 24** Kinetic energy of micrometeoroids impacting M1. The black line depicts the chosen velocity of 35 km/s for all particle masses. The dotted cyan and magenta lines show the approximate upper and lower bounds of kinetic energy for the full range of micrometeoroid velocities (~5 - 70 km/s).

### 3.3.3 Deformation of the Surface

Now that we understand how often we anticipate impacts for any given size, we need to understand how these impacts may deform the surface. Since M1 is an inflatable, deformation of the surface will involve compression of the gas. We approximate the initial shape of M1 as a sphere. We treat the gas as ideal, assume the temperature and number of moles remain constant during the interaction, and assume all kinetic energy of the particle goes into compression of the gas. Fig. 24 shows the kinetic energy of a particle with mass m striking M1 for various velocities. Equating the particle kinetic energy and the work done on the gas gives

$$\frac{1}{2}mv^2 = \int \vec{F} \cdot d\vec{l} \tag{9}$$

The work done on a gas under isothermal conditions can be expressed in terms of the volume change as

$$\int \vec{F} \cdot d\vec{l} = -nRT \ln\left(\frac{V_f}{V_i}\right) \tag{10}$$

where n is the number moles of the gas, R is the gas constant, T is the temperature of the gas, and $V_f$ and $V_i$ are the final and initial volumes of the gas respectively. Substituting this into the previous equations results in

$$\frac{1}{2}mv^2 = -nRT \ln\left(\frac{V_f}{V_i}\right). \tag{11}$$

Since we are interested in understanding the deformation of the M1 shape, we rearrange the above equation to solve for the final volume

$$V_f = V_i e^{-mv^2/2nRT}. \tag{12}$$

We can quantify the deformation in shape by calculating the change in radius due to the impact. The resulting radius after compression will be:

$$R_f = \left(\frac{3V_f}{4\pi}\right)^{1/3}. \tag{13}$$

For a for M1's volume n = 1.6 mols. Fig. 25 shows the change in radius for an impactor with mass m and velocities of 35 (solid trace) and 70 (dashed traces) km/s. The three traces for each velocity correspond to different temperatures of M1. The horizontal dash-dot-dot trace is a change in radius of a part in a thousand, our canonical tolerance. This can occur for particle masses greater than $10^{-7}$ g for v=70 km/sec and $10^{-6}$ for v=35 km/sec. From Fig. 23, we see that particles of this mass are expected less than once per day, even with directional enhancement, and should not be a significant impact on science operations.

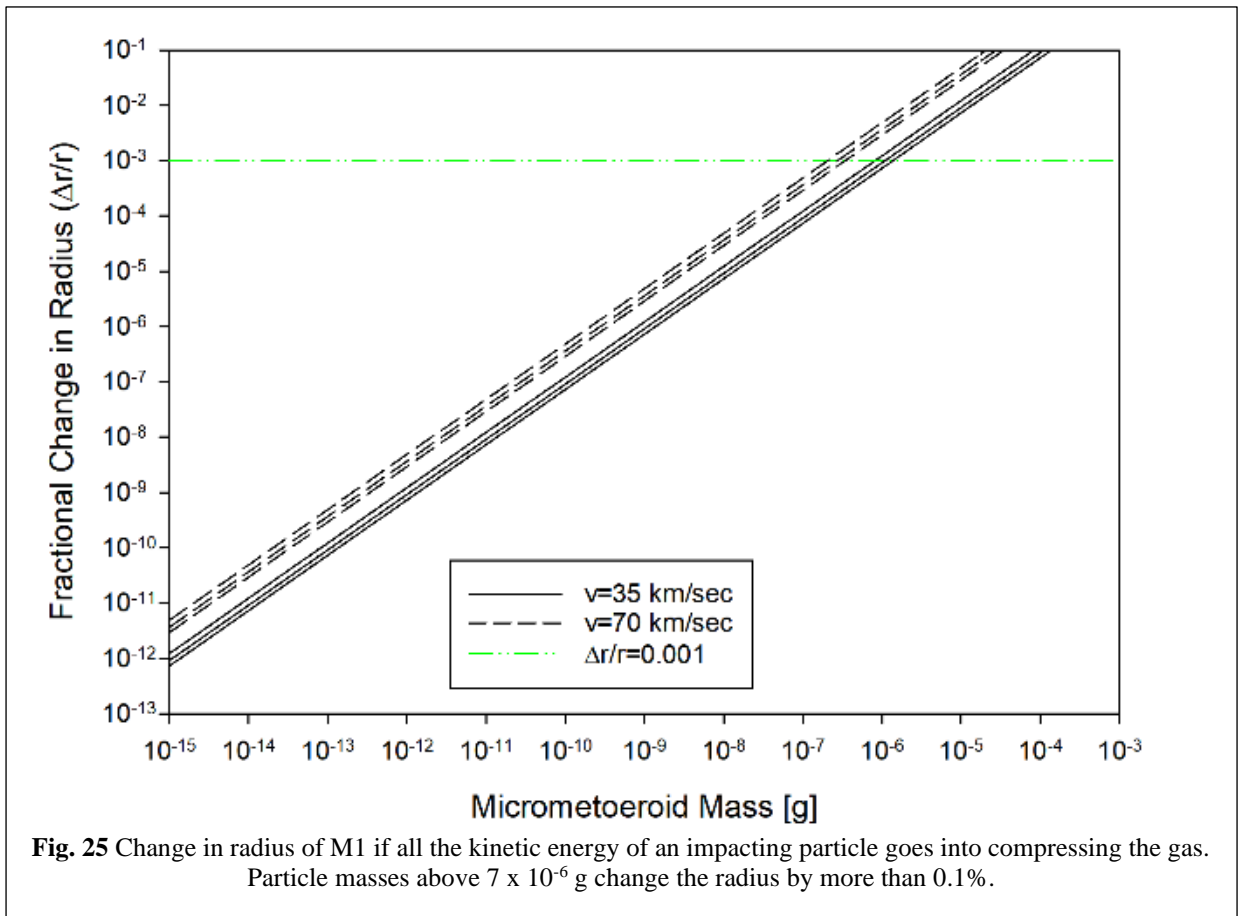

**Fig. 25** Change in radius of M1 if all the kinetic energy of an impacting particle goes into compressing the gas. Particle masses above 7 x $10^{-6}$ g change the radius by more than 0.1%.

We can compare the radius displacement to the behavior of a spring to understand how the rigidity of M1 compares to other telescopes. We equate the work done on a spring to compress it a distance $\Delta R$ to the work done to compress the gas

$$\frac{1}{2}k\left(\Delta R\right)^2 = -nRT\ln\left(\frac{V_f}{V_i}\right) \qquad\qquad (14)$$

where k corresponds to the spring stiffness. Solving the above equation for k gives:

$$k = -\frac{2nRT}{\left(\Delta R\right)^2}\ln\left(\frac{V_f}{V_i}\right). \qquad\qquad (15)$$

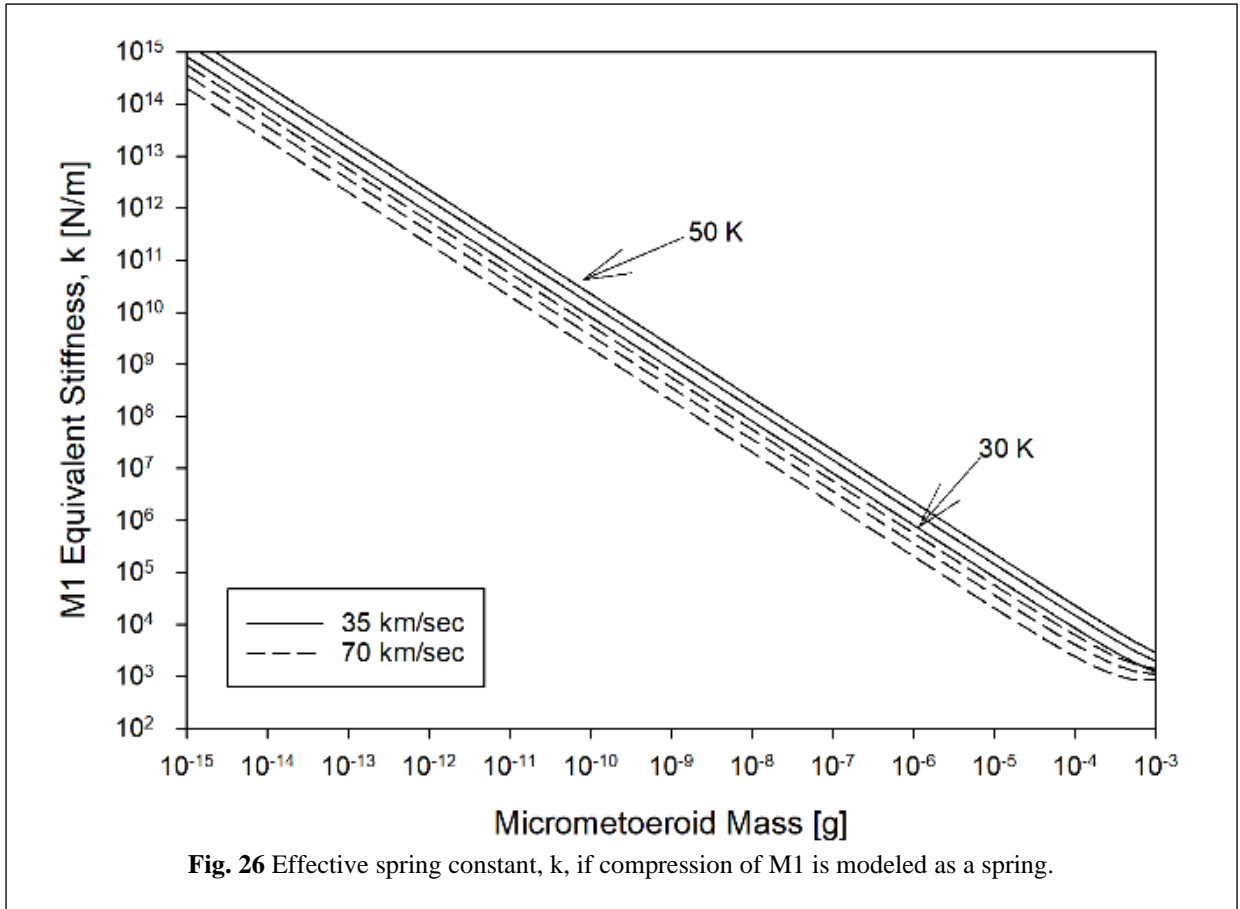

**Fig. 26** Effective spring constant, k, if compression of M1 is modeled as a spring.

Fig. 26 shows M1's equivalent stiffness to different particle masses. For comparison, mirror elements of the JWST have a stiffness range of $10^5$-$10^7$ N/m. M1 exhibits comparable stiffness for particles with masses up to $10^{-5}$ g, after which the effective spring constant is below currently used telescopes.

We can also think about the compression effect in terms of pressure. Since we have set the number of moles and temperature to be constant, the final pressure can be expressed as:

$$P_f = \frac{P_i V_i}{V_f}.$$
(16)

Fig. 27 shows the change in pressure due to a particle impact of a given mass. The red dotted line indicates the pressure tolerance of M1, $0.001P_0$. The pressure tolerance is exceeded for particle masses > 2 x $10^{-6}$ g for v=70 km/sec and 8 x $10^{-5}$ g for v=35 km/sec, for these particles return periods are longer than a day, meaning no serious impact to optical performance.

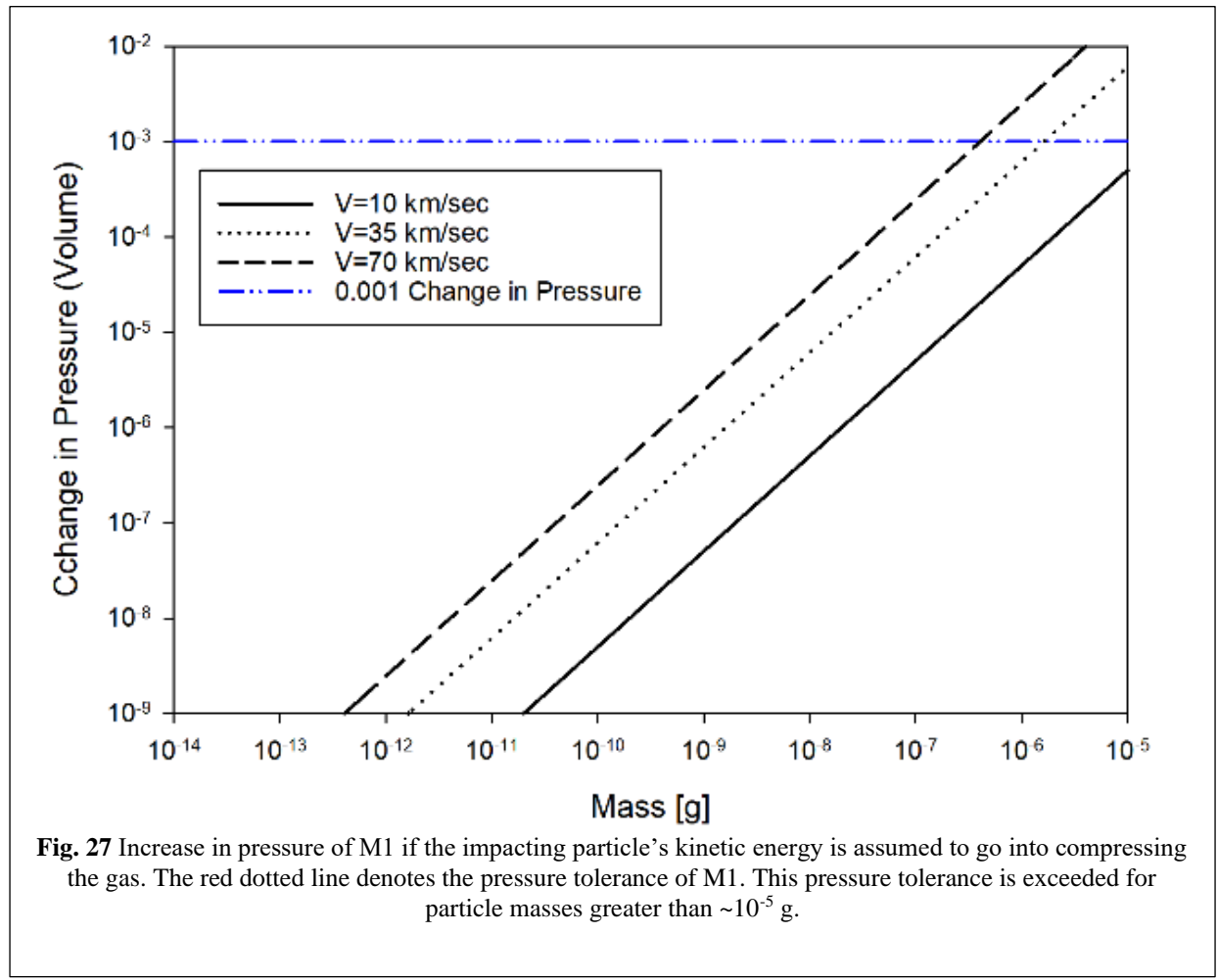

**Fig. 27** Increase in pressure of M1 if the impacting particle's kinetic energy is assumed to go into compressing the gas. The red dotted line denotes the pressure tolerance of M1. This pressure tolerance is exceeded for particle masses greater than ~$10^{-5}$ g.

### 3.3.4 Optical Impact of Punctures from Micrometeoroids

Section 3.3.3 describes potential disturbances to M1 from non-penetrating or elastic collisions. The current section deals with penetrating, non-elastic collisions which occurs when the particle is above the ballistic limit. The aftermath of such collisions are wounds in M1. These wounds are interruptions in the optical surface and the inflatant gas containment, the former are optical imperfection and the latter sources of gas loss. The holes will not impact optical performance. The holes in the reflective surface appear as small reflective losses and look like particles obscuring the surface. At end of life, this fractional area of the holes is $\sim 10^{-4}$, somewhere between MIL-STD-1246 Class 70 and 300, depending on slope, a very clean optical surface in any regard. [39] The analysis of gas loses from these holes is analyzed in 3.4.3.

### 3.4 Inflatant Longevity

This section examines the reasons for the consumption of inflatant, one two life-limiting consumables on spacecraft flight system. [2,40] The following discussion of how the total inflatant needs of M1 are calculated is intended to be a bounding argument and not a precise inflatant usage calculation. Namely, the goal of our present argument is to show that an inflatant reservoir will last at least a given years, at least 5 and out to 10 years, not that it will be exhausted at a specific hour.

The four sources of inflatant loss are, permeation through the membrane, effusion through MM induced holes, venting of excess gas when moving from a cold attitude to a hot one and valve and fitting losses. Each one will be examined in turn and at the end of this section the lifetime will be examined collectively under a variety of parameter arrangement to demonstrate the robustness of the longevity of the inflatant inventory.

### 3.4.1 Permeation of the Membrane

The permeability of gasses through polymeric material varies widely as a function of the permeant gas species, the membrane material and especially temperature. [41,42] Here we consider the case of helium permeating through a polyimide membrane. Data on the permeability characteristics of noble gasses through Kapton® film can be found in the work of Schowalter et al. [42] In particular, test results for the permeability of argon is provided at temperatures of 22 °C and 50 °C.

The permeation rate is given by

$$Q = K \frac{A}{d} \Delta P \tag{17}$$

where Q is the number flow rate, K is the permeability, A is the surface area, d is the membrane thickness and $\Delta P$ is the pressure differential of the permeant gas across the membrane (here simply the gas pressure within the reflector). The temperature dependence of the permeability is given by

$$K \propto \exp\left(-\frac{E_K}{k_B T}\right) \tag{18}$$

where $E_K$ is the energy of permeation.

The permeability determined by Schowalter et al [48] for a temperature of 22°C is

$$K = 1.5 \times 10^{-11} \text{ cm}^3 @ \text{ STP mm/(Torr cm}^2 \text{ sec)}.$$

For a pressure of 5.1 Pa, a membrane thickness of 0.5 mils (0.0127 mm) and a membrane area of 350.0 m² (corresponding to a physically inflated diameter of 15 m), the mass loss rate through both faces of the SALTUS M1 would be ~$1.6 \times 10^{-3}$ kg/year, or 1.6 g/yr, at the assumed gas temperature of 45 K. It is concluded that the gas loss via permeation is insignificant.

### 3.4.2 Micrometeoroid Induced Losses

The holes created by MM are inevitable, so we must mitigate against the gas loss through a sufficiently large reservoir of replacement inflatant. To understand the problem, we first derive an expression of how M1 will lose mass with time. With the initial result we determine the pressure corrections needed to maintain M1's shape (constant volume) over short and long time periods. Later in this section we will discuss the formation of penetration wounds in M1 to determine the rate of mass loss over mission lifetime.

### 3.4.2.1 Determination of Effusion Rate from M1

If a container of volume V has N molecules with mass m inside it at temperature T, their velocities are given by the Maxwellian distribution. The average number of collisions J over an area A, of a gas against a container wall is found by integrating over angle and velocity and is a classic textbook problem in the Kinetic Theory of Gasses.[43]. J is given by

$$J = \frac{N}{4V}\sqrt{\frac{8k_B T}{\pi m}} \tag{19}$$

The ideal gas law is

$$PV = nRT \tag{20}$$

where n, the number of moles in the container is $n = N/N_A$, $N_A$ is Avogadro's number, P is the pressure. Solving (20) for N/V gives

$$\frac{N}{V} = \frac{PN_A}{RT}. \tag{21}$$

Substitution of (21) into (19) gives

$$J = \frac{PN_A}{RT} \sqrt{\frac{k_B T}{2\pi m}} \; . \tag{22}$$

Equation (22) simplifies to

$$J = P \sqrt{\frac{N_A^2 k_B T}{2\pi m R^2 T^2}} \; . \tag{23}$$

Recalling $R = N_A k_B$ gives

$$J = P \sqrt{\frac{1}{2\pi m k_B T}} \; . \tag{24}$$

J is in collisions/area/time so to get the rate of mass lost, $\dot{m} = \frac{dm}{dt}$, due to a hole of area A, J is multiplied by A and the mass per molecule m resulting in

$$\dot{m} = JAm = -PA \sqrt{\frac{M}{2\pi N_A k_B T}} \tag{25}$$

where, A is the area of the hole, M is the molar mass, $k_B$ is Boltzmann's constant, T is the temperature and $m = M/N_A$. The total mass contained in M1 can be described in terms of the moles, n,

$$m = nM \; . \tag{26}$$

Substitution of (26) into (25) gives

$$\dot{n}M = -PA \sqrt{\frac{M}{2\pi N k_B T}} \; . \tag{27}$$

Cancelling out the factor of M in (27) and making the time derivative explicit results in

$$\frac{dn}{dt} = -PA \sqrt{\frac{1}{2\pi N k_B M T}} \; . \tag{28}$$

We know that the total area A of the holes caused by the MM environment is essentially a linear function of the elapsed mission time, S, which is

$$A = \alpha S \tag{29}$$

where A is the time dependent area lost to penetrations and $\alpha$ is the area lost per time. Substitution of (29) in (28) yields

$$\frac{dn}{dt} = -P\alpha S \sqrt{\frac{1}{2\pi N k_B M T}} \ . \tag{30}$$

Using the ideal gas law, we can substitute for pressure in terms of the other gas variables giving

$$\frac{dn}{dt} = -\frac{nRT\alpha S}{V} \sqrt{\frac{1}{2\pi N k_B M T}} \ . \tag{31}$$

Equation (31) simplifies to

$$\frac{dn}{dt} = -n \frac{R\alpha S}{V} \sqrt{\frac{T}{2\pi N k_B M}} \ . \tag{32}$$

It is important to note here that the timescale in which we are operating in (time between pressure corrections), there will be little change in S, V, and T. Therefore, while S, V, and T are functions of time in the long term, we treat them as constant for the purposes of solving (32) for the short time behavior. Equation (32) is a separable differential equation with constant coefficients and its solution is

$$n(t) = n(0) e^{-t \frac{R\alpha S}{V} \sqrt{\frac{T}{2\pi N k_B M}}} \ . \tag{33}$$

The characteristic time $\tau$ of this exponential decay is

$$\tau = \frac{V}{R\alpha S} \sqrt{\frac{2\pi N k_B M}{T}} \ . \tag{34}$$

A key point in our analysis of (34) is the appearance of the M1 elapsed mission time, αS, in the denominator. This means that the time to sense and correct any deficits in pressure becomes shorter as the mission goes on and more area is opened up. Equation (33) can be rewritten using (34) resulting in

$$n(t) = n(0) e^{-\frac{t}{\tau}}. \tag{35}$$

A key operating parameter is the time for the gas to leak to the point where P is out of tolerance, namely P(t) = P - δP. To determine the time between injections of gas to maintain pressure we write δP(t) as

$$\delta P(t) = \frac{n(0)RT}{V} - \frac{n(t)RT}{V}. \tag{36}$$

Factoring out common terms and using (35) results in

$$\delta P(t) = \frac{n(0)RT}{V}\left(1 - e^{-\frac{t}{\tau}}\right). \tag{37}$$

Equation (37) can also be written

$$\delta P(t) = P(0)\left(1 - e^{-\frac{t}{\tau}}\right). \tag{38}$$

Solving (38) for t gives

$$t = \tau \ln\left(\frac{1}{1 - \frac{\delta P}{P(0)}}\right). \tag{39}$$

Substitution of the definition of τ from (34) into (39) gives

$$t = \frac{V}{R\alpha S} \sqrt{\frac{2\pi N k_B M}{T}} \ln\left(\frac{1}{1 - \dfrac{\delta P}{P(0)}}\right).$$ (40)

Since δP<<P, (40) can be linearized resulting in

$$t = \frac{\sqrt{2\pi N k_B}}{R} \frac{1}{S} \frac{V}{\alpha} \sqrt{\frac{M}{T}} \frac{\delta P}{P(0)}.$$ (41)

The timescale for adding gas to M1 will be determined by design choices for mission lifetime (S), the geometric properties of M1 which define the volume, V. the environment which defines α, inflatant species (M), our pointing attitude (T), and error budget allocation (δP/$P_0$).

We now turn our attention to calculating the total gas needed over the mission lifetime, S. We previously defined the cumulative hole area due to impacts as a linear function of time, (29). We expand (29) into detailed terms as

$$A(t) = A_{A1} t \int \phi(w) \pi \left(\frac{\kappa w}{2}\right)^2 dw$$ (42)

where κ is a constant that relates the size of the hole to the size of the particle, $A_{M1}$ is the surface area of M1, t is time, and the φ(w) represents a function describing the micrometeoroid flux as a function of size (diameter). Equation (42) assumes that the hole made in the membrane is circular. Equation (42) is rearranged to give

$$A(t) = t\left(A_{M1}\kappa^2\right)\frac{\pi}{4} \int \phi(w) w^2 dw.$$ (43)

In (43), the time dependence is the first term on the right-hand side, the parenthetical term is a design dependent term, the area and the size of the damage hole (assumed constant with size and energy), is a numerical term and the integral term which is dependent on the natural environment. Associating the constant term in (43) as α, we can write the time rate of mass loss as

$$\frac{dm}{dt} = -tP\left(A_{A1}\kappa^2\right)\frac{\pi}{4}\sqrt{\frac{M}{2\pi Nk_BT}}\int \phi(w)w^2dw .$$ (44)

As complicated as (44) looks it is very simple, namely

$$\frac{dm}{dt} = -tQ$$ (45)

where Q is given

$$Q = P\left(A_{A1}\kappa^2\right)\frac{\pi}{4}\sqrt{\frac{M}{2\pi Nk_BT}}\int \phi(w)w^2dw$$ (46)

with a bounding lower value is used for T. In (46) all of the terms except $\kappa$ and $\int \phi(w)w^2dw$ are known and constant. The impact of a collision, the magnification the wound caused by a particle of size d, enters Q as $\kappa^2$ can be determined by hyper-velocity testing (see later discussion). The integral term is a property of the environment. The total mass needed for the mission is the mass needed to replace the total mass lost given by

$$m_{gas} = \int_0^S Qtdt .$$ (47)

Which trivially integrates to

$$m_{gas} = \frac{Q}{2}S^2 .$$ (48)

Equation (48) clearly shows the inflatant gas mass is proportional to the square of the mission length, $S^2$.

### 3.4.2.2 Area of M1 Lost to Micrometeoroid Penetration

When a MM exceeds the ballistic limit and penetrates a membrane such as M1, a wound results on both layers of the lenticular. As described at the top of this section, we seek a bounding analysis of inflatant consumption. The goal of this section is the bounding analysis of impactor wound area and its effects on inflatant consumption.

Equation (46) and its isomorphs show that the mass loss rate depends on many variables some are easy to comprehend such as pressure, temperature, gas species. The total area of penetration or total wound area is much harder to grasp with simple physical intuition. The evolution of the wound area over mission life must be estimated conservatively to make a robust assessment of the lifetime of an inflatable system necessary to the viability of the SALTUS mission concept.

We have recently reported on the propagation of MM in gossamer structures giving predictive model of fragmentation. [44] An even more recent report gives further experimental examination of fragmentation and penetrations of gossamer systems and allows us further insight into this complex problem. [45].

The impact of a MM on a series of gossamer layers is shown schematically in Fig. 28. The image shows the clean initial penetration and fragmentation of the impactor and the propagation of fragments creating exit wounds.

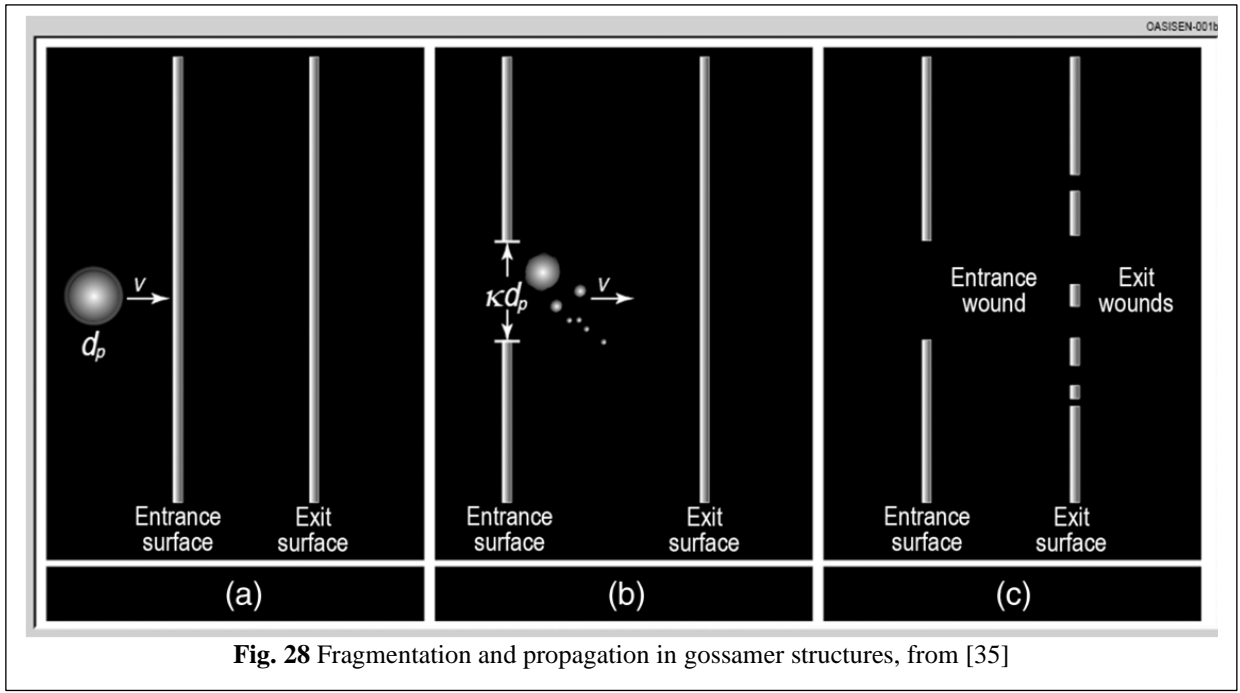

**Fig. 28** Fragmentation and propagation in gossamer structures, from [35]

Fig. 29 shows micrographs of recent hypervelocity tests and demonstrates the experimental evidence of fragmentation of a large impactor. The entrance wound is a clean circular wound, the shock from the collision shatters the impactor into fragments that follow a power law distribution. Each of the fragments from the collision with the first surface shatters at the second surface, each one into a power law, with the largest particle decreasing in size and the slope becoming steeper. This process of sequential fragmentation has been discussed recently.[44]

The recent tests carried out involved many layers of different materials and impactor velocities, to collect data to among other goals, further validate our recent propagation model. Examples of the typical results are shown in Fig. 30 and Fig. 31. Fig. 30, which is a test of typical ½ mil thick polyimide layers, this is same material used for the development of the propagation model.[44]

One of the materials recently tested and reported, was a ½ mil thick layer of a composite nature and significantly more damage resistant as shown in Fig. 31. [45] The distribution of fragments indicates smaller maximum particle size and steeper slopes indicative of much less wound area on the membrane.

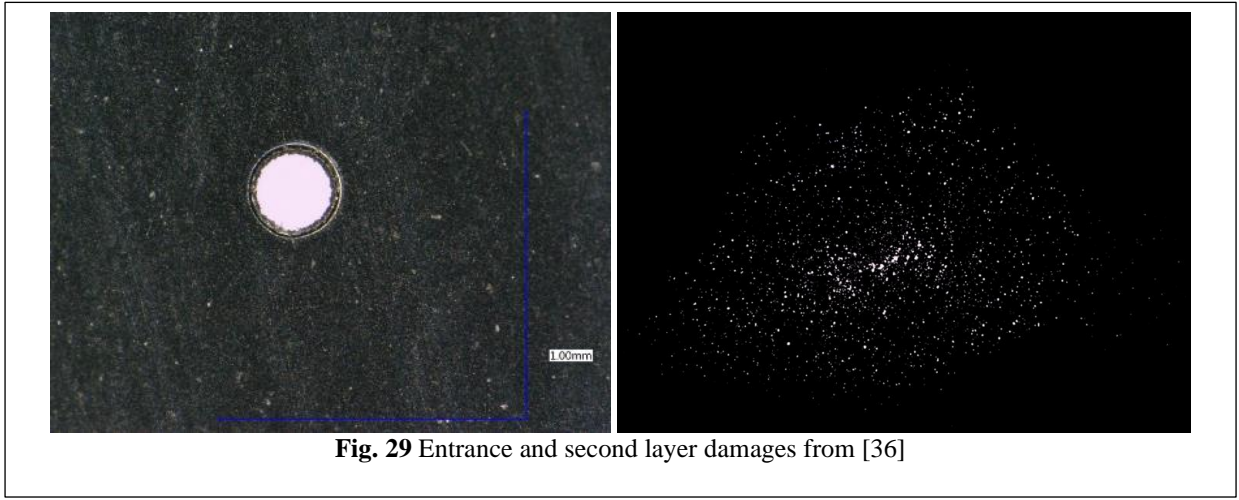

**Fig. 29** Entrance and second layer damages from [36]

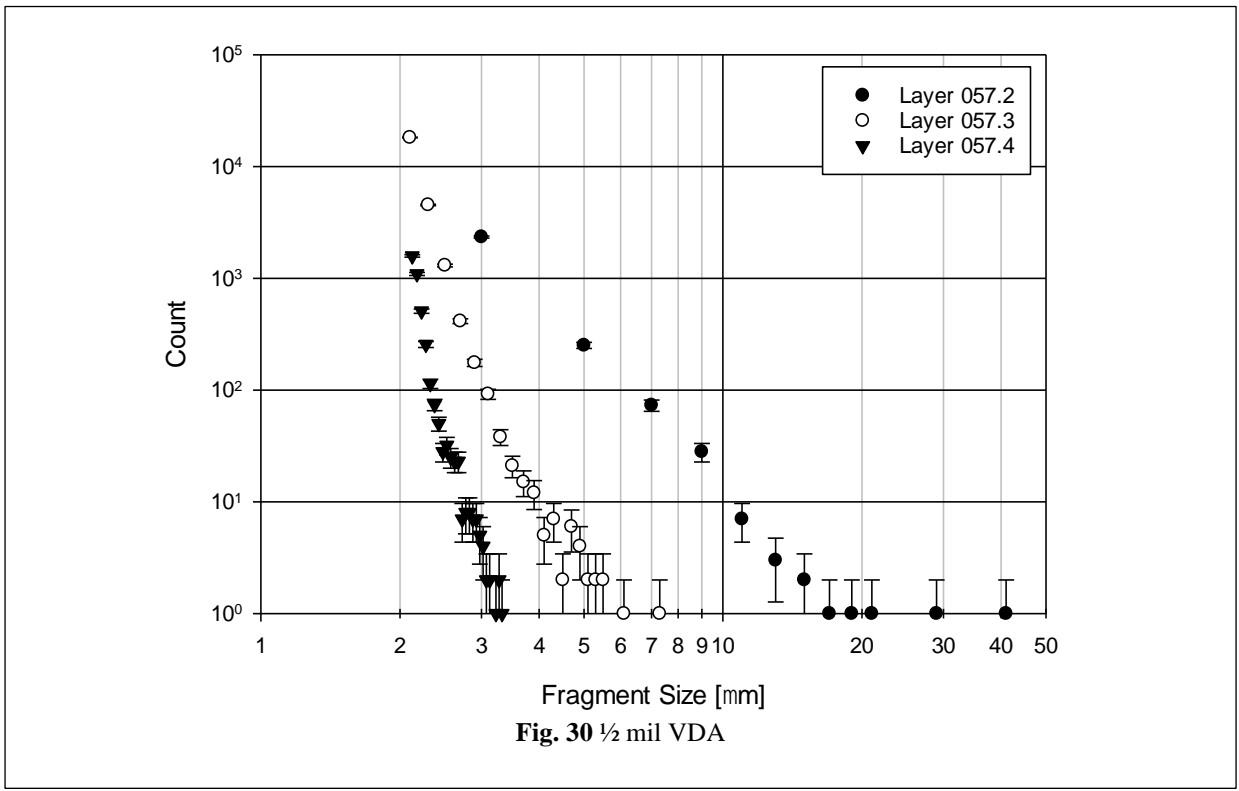

**Fig. 30** ½ mil VDA

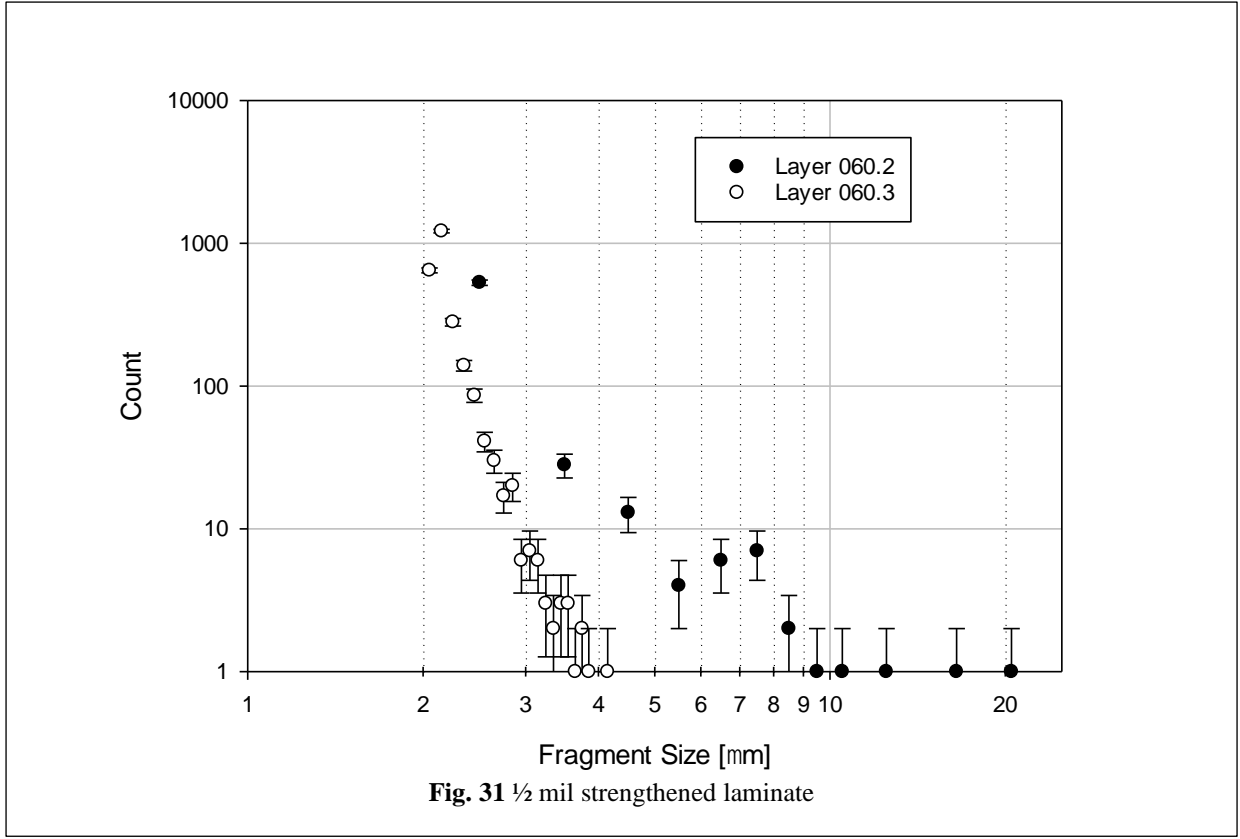

**Fig. 31** ½ mil strengthened laminate

One of the key results of our recent work [44], was the calculation of the specific area lost for various membrane thickness and numbers of layer for the entire flux of impactors. The estimate of specific area loss rate in units of cm$^2$/m$^2$/yr denoted $\Lambda$, for the baseline ½ mil polyimide material is the basis of our area estimation.

Creep can affect the determination of the overall wound area. A simple model of this effect is an increase in wound area with time. Let us call the increase in area with time $\gamma$ and furthermore assume that creep is constant in time. [46] The linear estimation provides a clear upper limit to this effect as creep at a given load and temperature explicitly a monotonically decreasing effect.

The linear approximation also makes the calculation easier to follow. Our simple (and bounding) model of hole area at time interval k including creep is written

$$A_K = a\big(1 + \gamma(k-1)\big) + a\big(1 + \gamma(k-2)\big) + \ldots + a \tag{49}$$

Where $A_k$ is the total area at the kth week, a is the area generated in a time interval. Equation (49) can be more compactly written as

$$A_k = \sum_{j=0}^{k-1} a\big(1 + j\gamma\big). \tag{50}$$

Equation (50) expands to

$$A_k = ka + \gamma a \sum_{j=0}^{k-1} j \tag{51}$$

Using the well-known formula for the summation of an arithmetic series (51) becomes

$$A_k = ak + \gamma \frac{k\big(a(k-1) + a\big)}{2} \tag{52}$$

with simplification we arrive at

$$A_k = ak + \gamma a \frac{k^2}{2}. \tag{53}$$

The term a is the rate of increase in hole area per time interval. For our current analysis, the time interval is a week and is written

$$a = 2A_{M1}\frac{\Lambda}{52} \tag{54}$$

where $A_{M1}$ is the physical area of M1, the factor of two is included as there are two membranes in M1. Substitution of (54) into (53) gives

$$A_k = A_{M1} \frac{\Lambda}{26} \left( k + \gamma \frac{k^2}{2} \right). \qquad (55)$$

The inclusion of the impact of creep (55) in the analysis that lead to (48) will introduce a term proportional to the cube of mission time, S. Clearly, if γ large, meaning the second term in the parenthetical expression in (creep55) is of order k then the entire term $A_k$, will be larger and sensitive to this term. At the low temperatures M1 will be operating, creep is expected to be nil [46] and can be easily tested early in the program before making a choice of the flight membrane material.

### 3.4.3 Cold to Hot Pointing Losses

The temperature of M1 will change due to differences in solar illumination as the spacecraft attitude changes over the FOR. Since we want the shape of M1 to remain static, we will need to design for adjustments in the gas during these events. We calculate the difference in moles of inflatant needed to maintain M1 at constant pressure and volume as the gas temperature changes from hot to cold, or vice versa. The ideal gas law under these conditions gives

$$nT = \frac{PV}{R} = \text{constant} \qquad (56)$$

where n is the number of moles, T is the temperature, P is the pressure, V is the volume, and R is the gas constant. We can then describe the change in our initial conditions to our final conditions as

$$n_i T_i = n_f T_f . \qquad (57)$$

Rearrangement of (57) gives

$$n_f = \frac{n_i T_i}{T_f}.$$ (58)

We can define $n_f$ in terms of the adjustment or change in number of moles needed by

$$n_f = n_i + \Delta n.$$ (59)

Substitution of (59) into (58) gives

$$\frac{n_i T_i}{T_f} = n_i + \Delta n.$$ (60)

Solution of (60) for results in

$$\Delta n = \frac{n_i T_i}{T_f} - n_i,$$ (61)

$$\Delta n = \frac{n_i T_i}{T_f} - \frac{n_i T_f}{T_f},$$ (62)

finally

$$\Delta n = \frac{n_i \left( T_i - T_f \right)}{T_f}.$$ (63)

From (63) we can see that when SALTUS points from a cold to hot attitude $T_f > T_i$ then $\Delta n < 0$ and gas must be released from M1.

The flux and impact velocity of the MM will depend on cosθ, where θ is the angle between the surface element's normal and the sun vector. The sun vector in the body coordinate frame is given by

$$\hat{s} = \langle \cos\beta \cos\gamma, -\sin\gamma, \sin\beta \rangle \qquad (64)$$

where $\beta$ is the solar pitch and $\gamma$ is the solar roll. We can then calculate $\cos\theta$ using the dot product of the sun vector and sunshield normal, (63) and (64) to give

$$\cos\theta = \hat{n} \cdot \hat{s} . \qquad (65)$$

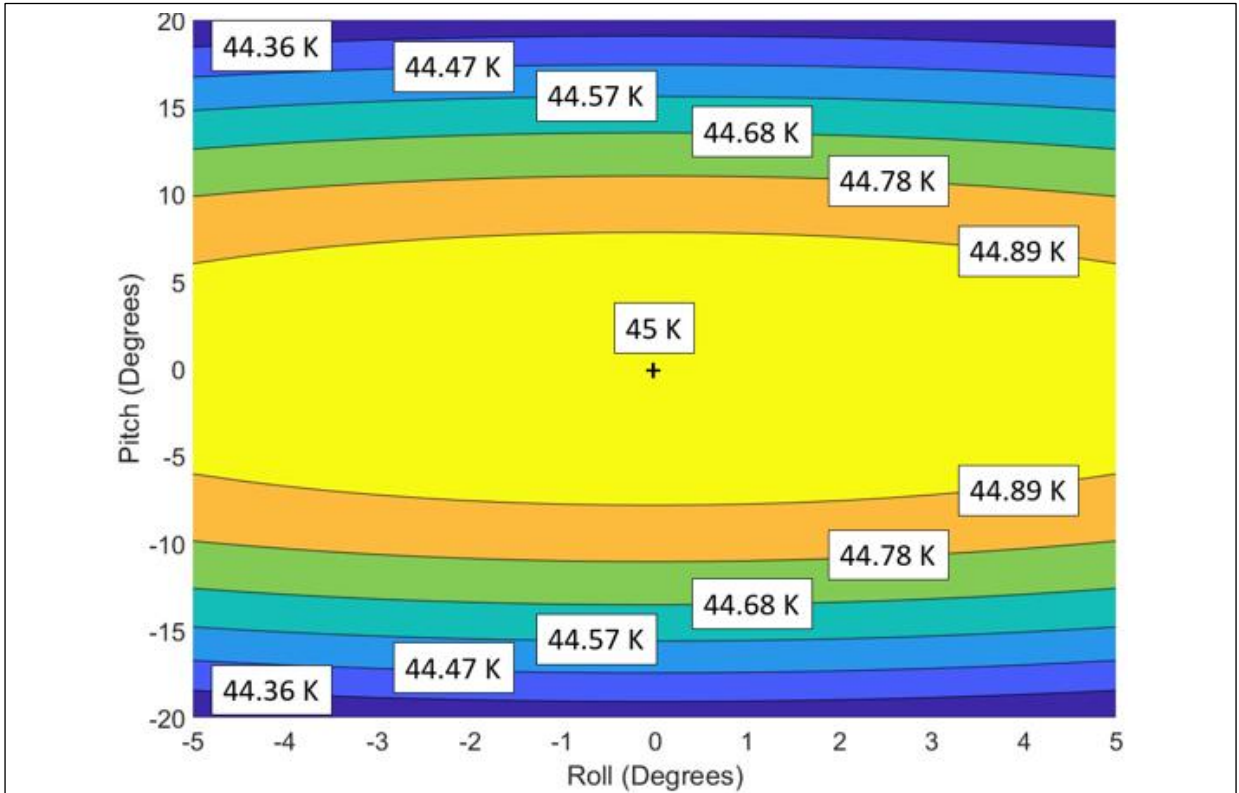

**Fig. 33** Map of Peak M1 Temperature

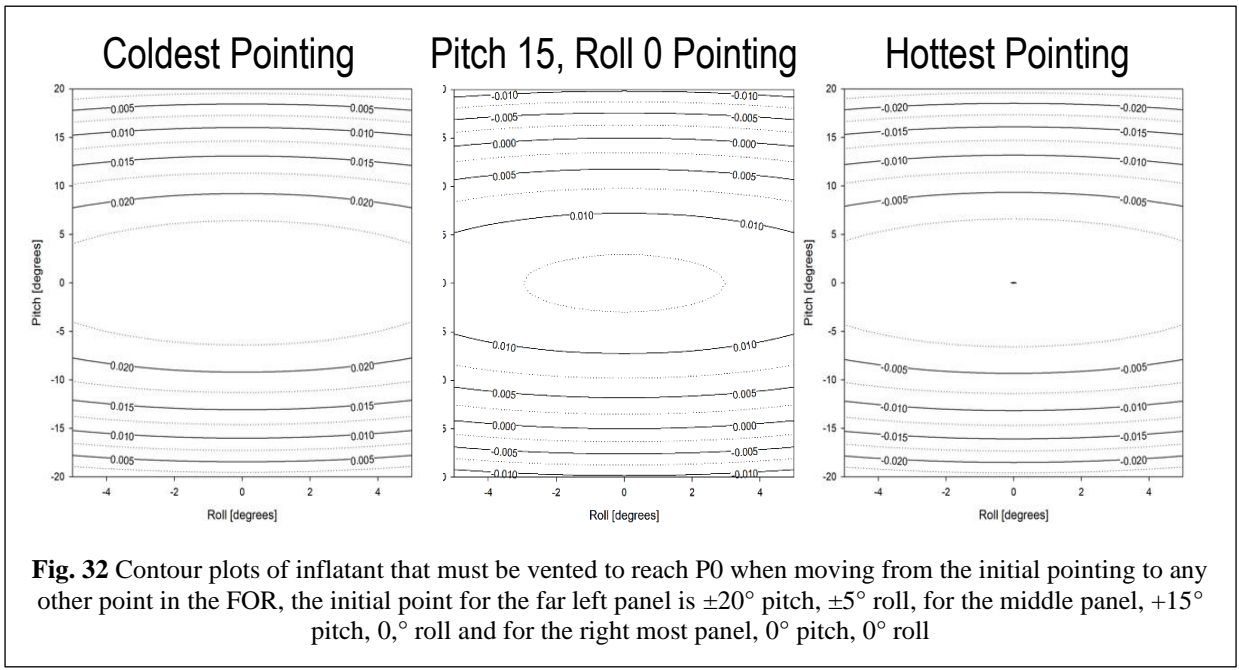

**Fig. 32** Contour plots of inflatant that must be vented to reach P0 when moving from the initial pointing to any other point in the FOR, the initial point for the far left panel is ±20° pitch, ±5° roll, for the middle panel, +15° pitch, 0,° roll and for the right most panel, 0° pitch, 0° roll

Equation (63) shows that Δn is negative if $T_f > T_i$, a cold to hot pointing change. Fig. 32 evaluates (63) over the full range of possible pointings (temperatures) given an initial pointing (temperature). The three initial pointing conditions are the three panels in Fig 35, the coldest pointing, ±20° pitch, ±5° roll, an attitude near the middle of the FOR, +15° pitch, 0° roll and the hottest pointing, 0° pitch, 0° roll. For the coldest pointing, the leftmost panel, every other pointing attitude requires venting of inflatant, a maximum of 0.025 moles. For the middle pointing, a move to a lower pitch value requires venting, maximum of ~0.017 moles, and moves to greater pitch angles require no venting. When the initial pointing is at the center of the FOR, all attitudes are cooler and require no venting upon repointing.

### 3.4.4 Valve Losses

Valve loses both internal and external are given by the manufacturers and are standard measures of performance. For a conservative estimate we assume that all valves leak for the entire mission, and we assume the amount of mass loss is the sum of the internal and external leaks. This is truly conservative as the internal leak will end up in M1 and simply lower the amount of gas the ICS injects. We, therefore, ignore this complication for this bounding estimate.

### 3.4.5 Inflatant Lifetime Net Assessment

As described above the net use of inflatant comes from four main sources, each with their own complications. In this section we report on the sum in a maximum use case, which in some sense is not completely physical or case consistent but is constructed to show that even in this worst case, the inflatant will last 5 years and meet mission life requirements. Under more realistic conditions the gaseous helium reserve will last all 10 years of the expected extended mission. For the purposes of the current analysis we define three cases, the least consumptive case (LCC), the current best estimate (CBE) (of conditions) and the most consumptive case (MCC). These cases will help bound inflatant usage and show that the baseline design will last at least 5 years required under the most pessimistic case and is very likely to last all 10 years of the potential extended mission.  The most optimistic conditions include easily imaging increased shielding, l included with the stiffening mesh in the truss and protecting the back of M1. Our analysis shows how effective it might be, but this additional shieling is not part of the SALTUS design at the time of this writing.  The LCC temperature of 25 K was chosen for its better sensitivity and that impact on science, rather than for lifetime only. Later in this section we will relax that assumption, see Fig. 36. For conditions we assume that SALTUS incurs the maximal venting going from cold to hot attitudes. The rate of the venting is determined by the observation period for each set of conditions shown in Table 2.  We assume that the attitude goes from hot to cold and then cold to hot, so this venting penalty is incurred every other observation, with the period determined by twice the observation length.

**Table 2** Definition of operating conditions, most optimistic, best estimate, and most consumptive

|  | Least Consumptive Case (LCC) | Current Best Estimate (CBE) | Most Consumptive Case (MCC) |
|---|---|---|---|
| **Temperature [K]** | 25 | 37 | 25 |
| **Λ [cm²/m²/yr]** | $1.6 \times 10^{-6}$ | $1.6 \times 10^{-6}$ | $3.2 \times 10^{-6}$ |
| **Creep (/week)** | $10^{-6}$ | $10^{-6}$ | 0.001 |
| **Shielding** | 0.6 | 0.15 | 0 |
| **Obs length (h)** | 4 | 4 | 2 |

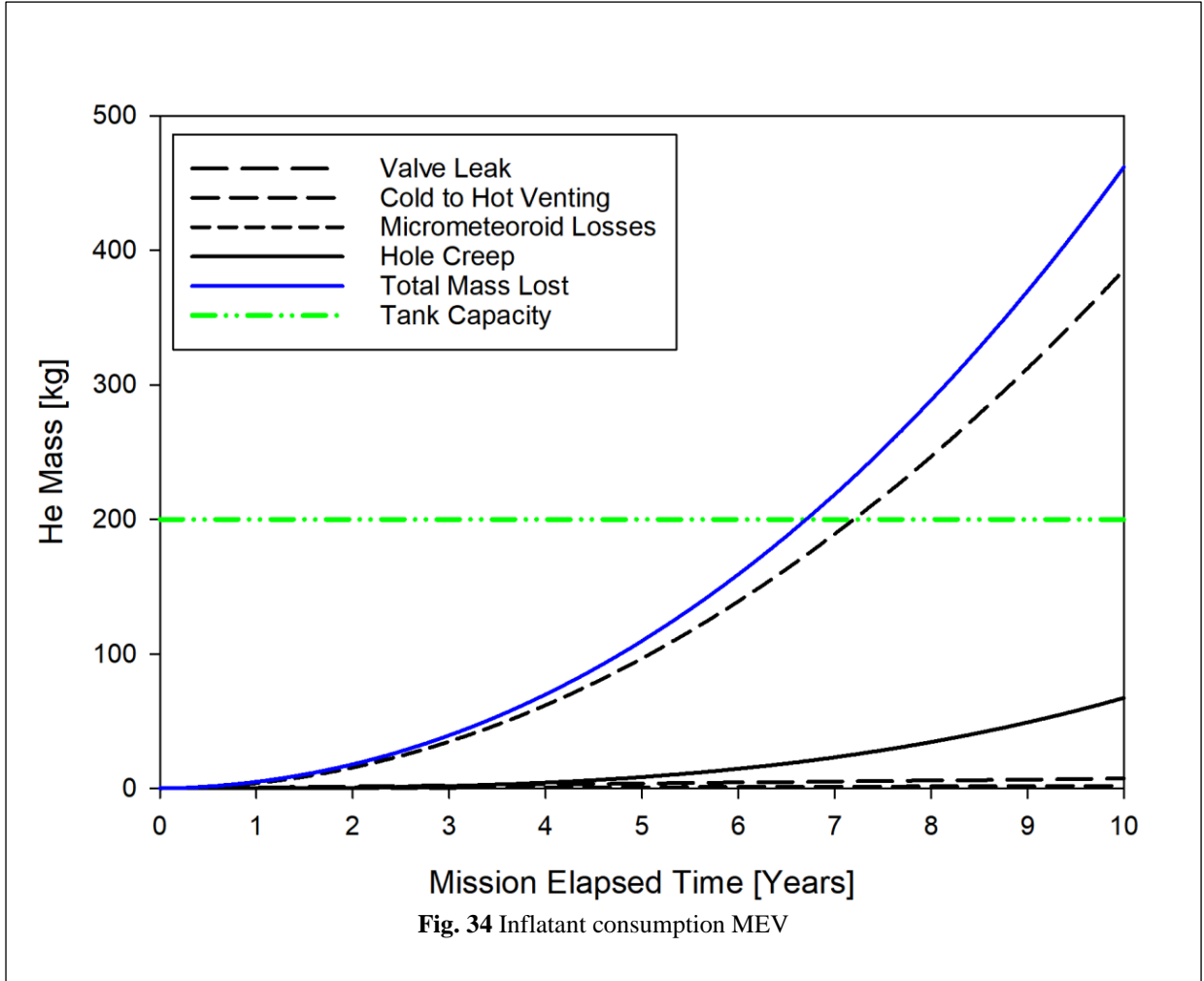

**Fig. 34** Inflatant consumption MEV

Fig. 34 shows the amount of inflatant for the four consumptive terms under the MCC conditions. The largest consumption is from micrometeoroid holes, followed by the extra consumption caused by creep. As noted, before, at the operating temperatures, this is quite unphysical and is included to make a true MCC or worst-case analysis. The cold to hot venting and valve losses are very small, but visible. Fig. 34 clearly shows that the inflatant will last 5 years and will be exhausted about 9 years, well past requirements.

Fig. 35 shows the time dependent usage of the inflatant in the CBE conditions. The order of consumption is the same, save for creep which effectively nil. Under this more realistic set of conditions the inflatant will last all 10 years.

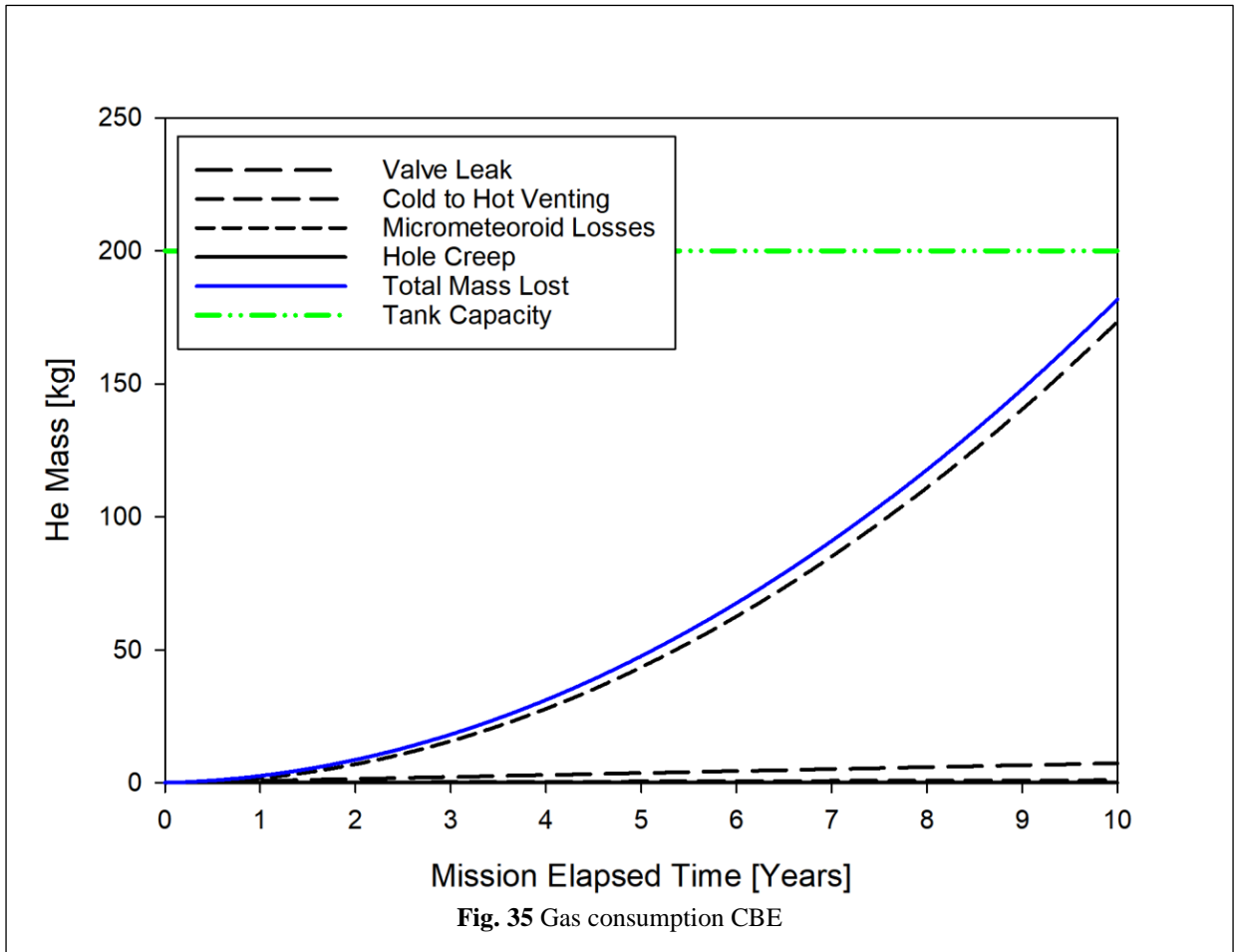

**Fig. 35** Gas consumption CBE

As the reader can appreciate from such results as (46) that inflatant consumption is a complicated performance space depending on many parameters. To give the reader more insight consider Fig. 36, which shows the necessary amount of gas for the required 5-year mission and the possible extension to 10 years for the LCC, CBE and MCC cases. As one of the most influential variables in gas consumption is T, T is varied from 25-45K. It is clear from the figure that in all cases, the minimum required lifetime of 5 years is met and exceeded. Only in the MCC case is the 10-year lifetime not met, and then not by a large margin.

To give further insight into the endurance of the inflatant inventory the lifetime as a function of M1 temperature was calculated. For LCC and all temperatures considered, the lifetime of the helium was longer than 10 years. The results for the CBE and MCC conditions have been explicitly calculated and plotted in Fig. 37. This figure clearly shows that for any M1 temperature, the minimum requirement of 5 years is met, and that under CBE conditions the inflatant is likely to last at least 9 years and for T>37 K a 10 year or slightly longer will result.

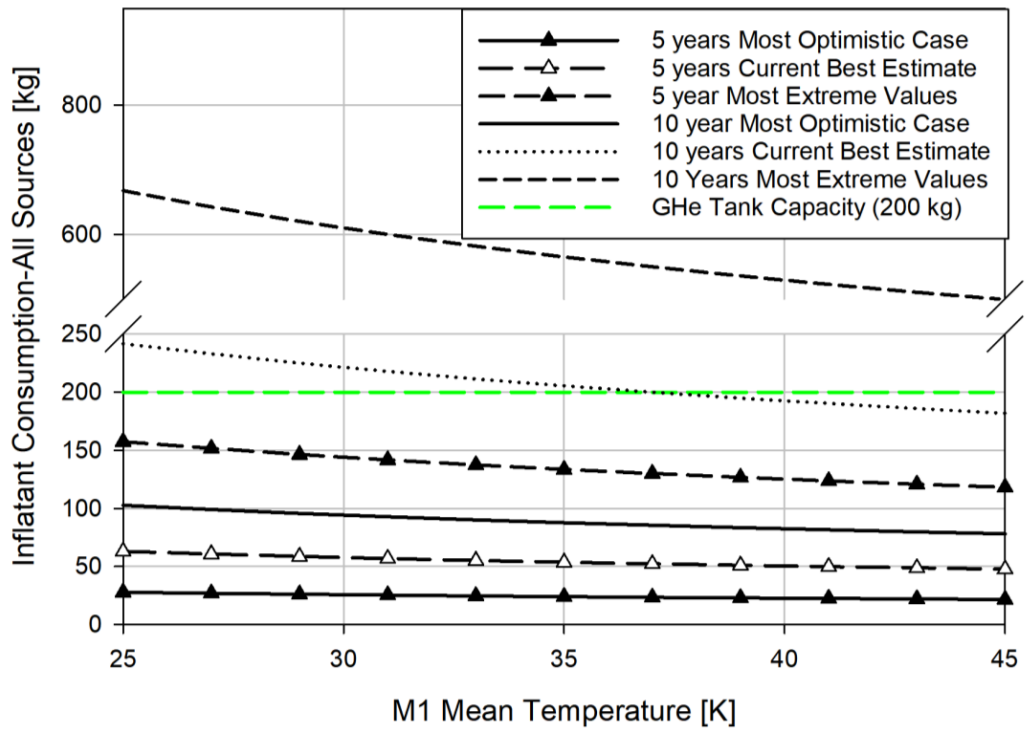

**Fig. 36** Inflatant requirements for LCC, CBE and MCC cases over M1 temperatures from 25 to 45 K

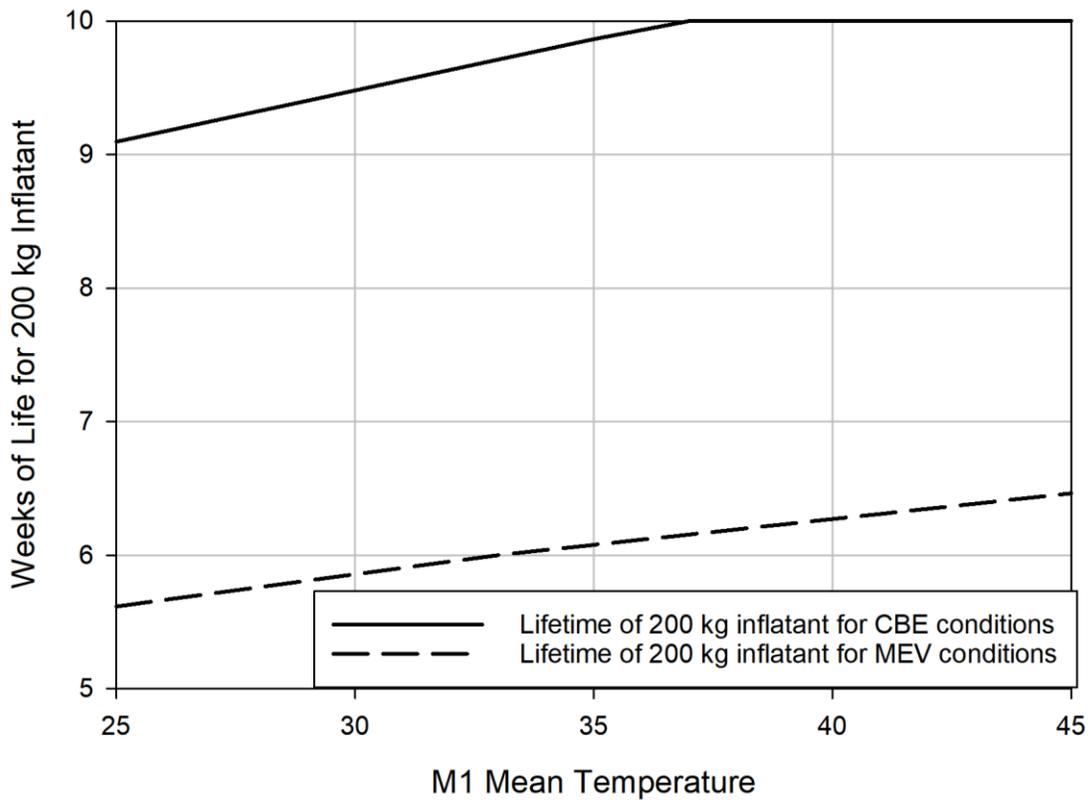

**Fig. 37** GHe lifetime for CBE and MCC conditions over M1 temperature from 25 to 45K

## 4 Planned Development Activities

**Table 3** Tests on the SALTUS Inflatable Reflector  Articles

| Test | 8M EDU | 14M Qual | 14M Flight |
|------|--------|----------|------------|
| Surface measurement on support | Yes | Yes | Yes |
| Packaging & Deployment without AstroMesh | Yes | Yes | Yes |
| Inflation | Yes | Yes | Yes |
| Ascent-Vent | Yes, at Flight Condition | Yes, at Flight Condition | Yes, at Flight condition |
| Thermal Cycling | Yes | Yes | |
| Packaging & Deployment under vacuum | Yes, at NASA Glenn NATF | No | No |
| Packaging & Deployment | Yes | Yes | Yes |
| Random Vibration | Yes | Yes, at Qual Level | Yes, at Accep. Level |
| Surface Measurement | Yes | Yes | Yes |

Achieving TRL6 requires demonstration of performance of a relevant prototype in the proper environment. The chosen 8m article is the relevant prototype as it allows us to complete the necessary testing before PDR as required within program funding and cost constraints. To achieve TRL6 we must demonstrate that M1 can be constructed with the necessary precision, stowed, deployed and inflated and maintained in a relevant environment.  Thus, a series of tests shown in Table 3 are planned to verify the design and inform subsequent development articles. Precision of the M1's shape is determined by the tolerances on the cutting of the gores and their seaming and proportional to the size. To demonstrate sufficient accuracy at flight scale, the 8m development article will have gore and seam tolerances 8/14 of the flight values. We must demonstrate the requisite tolerances at 8m diameter that will give an acceptable 14 m M1. At 8m the article can be installed in the 9m "IRAD Reflector" a capital asset of Northrop Grumman, to demonstrate the ability to integrate, stow, deploy and survive deployment environments at the reflector/truss level. These deployment tests will allow multiple stow and deploy cycles and validate articulation margins and stowing procedures. Furthermore, a flight-like stow and environmental testing, will show that we will not crease M1. The deployment and inflation pf the development M1 will demonstrate the reflector to truss interface.

After testing at Northrop Grumman's Oxnard facility, the 8m reflector (M1 development article), the development ICS and optical metrology equipment go to NASA's Neil Armstrong Test Facility (NATF) in Ohio. At NATF the 8m test article will be integrated into a GSE support ring and inflated via the ICS (per Table 4) at atmospheric pressure and room temperature, to various pressures around the 5.1 Pa (gauge) and with several gasses (gravity sag) and the shape measured.  We will also tip the GSE interface ring at a few angles to provide a diversity of gravity environments. The 8m reflector will then be placed in the environmental chamber – this is the 3$^{rd}$ reason to select the 8m size, as it will fit in the chamber. The air will be removed from the chamber in steps allowing shape measurements at various atmospheric pressures. After reaching vacuum, the temperature will be lowered via the $LN_2$ shroud. A GSE heater plate will simulate the radiant load from the sunshield, the heater plate will have several independent zones so that the gradient across the test reflector can be varied. Data from these tests will validate the thermal and distortion models. Each of these test conditions provide a unique test environment in terms of temperature, pressure and gravity. Shape and other telemetry such as ICS sensor data will be compared to predictions and the shapes compared to an *a priori* established performance requirement. The data collected will also be used to update and validate the design models. At this point M1 will be at TRL6.

**Table 4** System Development Test Objectives: Test at NASA NATF with 8-m EDU M1.

| Test Objective | Test Approach | Model Validated | Use |
|---|---|---|---|
| Inflatable design process validation | Collect and trend optical metrology data of M1 from manufacture through end of test | Design codes (FAIM, FLATE) and flight FEM (system integrated model) | Design and on-orbit performance |
| Optical performance of M1 over temperature | Heater plates to adjust M1 temperature, vary inflation gasses to validate gravity model of M1 | System integrated model | Prediction of on-orbit performance |
| Demonstrate M1 shape over multiple inflation cycles | M1 shape data collected over periods of the ICS being active and then idle | Lifetime model of membrane behavior | Predicting and budgeting lifetime effects |
| Understand impact of ambient pressure on M1 inflated shape | Collect M1 metrology at room temperature and press and vacuum and low temperature | Design codes (FAIM, FLATE) and flight FEM (system integrated model) | Allows shape verification to occur at ambient |
| Validate creep models | M1 metrology data over the test including a high temperature test | Lifetime model of membrane behavior | Predicting and budgeting lifetime effects |
| Inflation control system demonstration | Use EM of ICS to actively control pressure, | ICS operating models | ICS System TRL maturity |
| Inflant consumption | Measure leak rates, from fittings and deliberate holes | Inflatant consumption model | System lifetime prediction |

These experimentally validated design models, for gravity, temperature and pressure and pressure control will be used to design the 14-m M1 units. M1 will be measured in various stressing environments to show compliance to requirements in the test environments that are consistent with flight performance. At this point M1 is at TRL7.

The existence of the set of validated models for all the relevant environments and environmental changes, allows the verification of flight M1 shape, via measurements in a factory environment.

## 5   Summary

We have presented the mission enabling element of SALTUS, its large inflatable primary reflector, M1, and the reasoning behind our confidence in being able to realize this game changing design. We have reminded the reader that inflatable systems are not a novel idea. The novelty in M1 is its accuracy (built on advances in manufacturing) and its longevity (estimated from extensive micrometeorite analysis and experiments). We have discussed the design process and solution of the problem of cutting and seaming two-dimensional gores to make an accurate near parabolic shape. We have also given analysis of how M1 will function in its environment and how the design mitigates these responses. The sufficiency of the planned inflatant inventory has also been demonstrated to be robust, meeting the minimum 5-year mission under worst case analysis. Finally, we give an overview of the specific development and testing planned for M1 to deliver a high-performance primary for SALTUS.

Taken together, this is why the time is right to bring M1 into operation, powering a revolution in Far-IR science and living up to the promise of the mission's name. We are ready to make this great leap.

*Conflict of Interest*

The authors declare no specific financial interests or other potential conflict of interest with the work presented in this report.

*Data Availability*

Data sharing is not applicable to this article, as no new data were created or analyzed.

*Funding*

Funding for the authors from Northrop Grumman, L'Garde and the University of Arizona was performed under respective internal funding. DA2 Ventures personnel was supported by funding from the University of Arizona.

*Acknowledgments*

The authors wish to acknowledge our internal funding and support for this work as well as the rest of the SALTUS team.

**Jonathan Arenberg** is currently the Chief Mission Architect for Science and Robotic Exploration at Northrop Grumman. His work experience includes the Chandra X-ray Observatory, co-invention of the starshade and the James Webb Space Telescope. He is the Chief Architect and Project Systems Engineer for SALTUS. He is an Associate Fellow of the AIAA and a Fellow of SPIE.

**Leon Harding** is a Senior Staff Engineer at Northrop Grumman. He received a PhD in astrophysics from the University of Galway in Ireland. Before Northrop Grumman, he held a Research Associate Professorship at the Virginia Tech National Security Institute and was the Assistant Director of the Mission Systems Division. He was a Technologist at the Jet Propulsion Laboratory, and a Scientist at Caltech. He is the Spacecraft Lead and Deputy Chief Observatory Architect for *SALTUS*.

**Bob Chang** is currently in the Business Development team of the Northrop Grumman Deployables Operating unit which is responsible for large space deployables such as booms and AstroMesh™ reflector assemblies. Prior to this current role, Bob spent 10 years as a mechanisms engineer with contributions to various programs such as GOES, Osiris-REX, SWOT, Parker Solar Probe, Bepi Colombo and JUICE.

**Steve Kuehn** is has worked at NGAS Deployables, formerly Astro Aerospace, since 1989 as a structural analyst. Starting in 1995, he has been involved in the development of the AstroMesh™ reflectors and in several AstroMesh™ flight programs. He obtained a BS in Mechanical

Engineering from California Polytechnic State University, and an MS in Mechanical Engineering from Stanford University.

**Dave Oberg** is a Program Manager with Northrop Grumman, in the Science and Robotic Exploration unit. In this position he works with scientists on initial concept development and then leads the team through design, production, and launch. He holds a Bachelors and Masters degree in Aeronautics and Astronautics from the Massachusetts Institute of Technology.

**Michaela N. Villarreal** is a physicist at Northrop Grumman currently working on navigation systems and sensors and was previously a systems engineer as part of their Science and Robotic Missions group. She received her bachelor's degree in planetary science from UC Berkeley and her MS degree and PhD from UCLA in geophysics and space physics. She was previously a science team affiliate of NASAs Dawn and Europa Clipper missions.

**Arthur Palisoc** is currently the Chief Technology Officer at L'Garde, Inc. His work experience includes the design and testing of the surface accuracy measurement system for the 14 m diameter inflatable antenna experiment reflector flown in 1996 on board the STS-77 Space Shuttle mission. He co-developed FAIM, the finite element analyzer code for inflatable membranes and used extensively in the characterization of the geometric nonlinear behavior of membrane shapes under load.

**Christopher K. Walker** is a Professor of Astronomy, Optical Sciences, Electrical & Computer Engineering, Aerospace & Mechanical Engineering, and Applied Mathematics at the University of Arizona since 1999. He has worked at TRW, Jet Propulsion Laboratory, was a Millikan Fellow in Physics at Caltech. He has led numerous NASA and NSF projects, authored/coauthored 130+ papers, and published two textbooks: "Terahertz Astronomy" and "Investigating Life in the Universe".

**Daewook Kim** is an associate professor of Optical Sciences and Astronomy at the University of Arizona. He has devoted his efforts to a multitude of space and ground-based large optical engineering projects over a broad range of wavelengths, ranging from radio to x-ray. He has written over 300 journal/conference papers and is an associate editor for Optics Express. He is an SPIE Fellow and was elected to the SPIE Board of Directors for the term 2024 to 2026.

**Zach Lung** is a Space Program Strategist, Business Development Manager, and Scheduling Lead at DA2 Ventures. He received his BS in business from Colorado State University in 2016. His professional background includes mission planning, proposal management, and technical writing for astrophysics, heliophysics, and ground-based Astronomy observatories, including work on the Giant Magellan Telescope (GMT) as part of the US Extremely Large Telescope Program (US-ELTP).

**Dave Lung** has over 40 years of aerospace industry experience in a wide variety of air and space technologies and systems. He is currently Founder/President of DA2 Ventures, a consulting firm specializing in tech transfer, program planning, capture management, and proposal management. His education includes BS Aerospace Engineering, University of Minnesota (1985), MS

Reliability Engineering, University of Arizona (1987), MBA University of Texas (1988), and post-graduate studies in Systems Engineering (USAF-AFIT). Officer, USAF (1984-1992).

**Caption List**

**Fig. 1** 3D rendered image of the SALTUS observatory, featuring an inflatable 14-m diameter off-axis primary mirror technology and its sunshield for radiative cooling below 45K (image credit University of Arizona).

**Fig. 2** Literature on Inflatable Space Telescopes from the past 40 years.

**Fig. 3** AstroMesh™ reflector and truss

**Fig. 4** (a) Solving the Inverse Problem. (b) Reflector made with pie shaped gores. (c) An inflatable reflector made with 24 Aluminized Kapton gores (image credit L'Garde).

**Fig. 5** Surface accuracy as a function of inflation pressure

**Fig. 6** Typical mesh for modeling (image credit L'Garde)

**Fig. 7.** Exploded view of rms surface versus pressure. In this figure, the error is relative to the minimum surface error.

**Fig. 8** Focal length as a function of inflation pressure

**Fig. 9** (a) 3m diameter inflatable lenticular showing its scalloped perimeter doubler edge support. (b) Fully inflated 1m diameter lenticular with inflatable torus support, diameter inflatable solar concentrator built by L'Garde for TransAstra Corporation in support of their *in-situ* resource utilization activities. (image credit L'Garde).

**Fig. 10** Thermo-formed Kapton membrane (image credit L'Garde).

**Fig. 11** 3m diameter gores laid out and bonded (image credit L'Garde)

**Fig. 12** (a) 7-meter diameter, F/D = 0.5 On-Axis Reflector Canopy. (b) IAE on orbit. (Figure a image credit L'Garde; Figure b image credit NASA)

**Fig. 13** A representation of the packaging of the inflatable lenticular. (A) Outer doubler shown (without inflatable). (B) Lenticular folded to the same diameter as available within the stowed AstroMesh truss. (C) The canopy and reflector are shown folded as cones whose projection on the plane is similar to the opening of the central "hole" in (B). The reflector and canopy are then Z-folded as depicted in (D) blue and green (image credit L'Garde).

**Fig. 14** Absorption cross section, Qbas, for candidate M1 contaminant species, aluminum, carbon and salt (NaCl)

**Fig. 15** Inflation Control System block diagram

**Fig. 16** Contours of layer 2 emissivity giving $T_{M1}$=45 K as determined by (5)

**Fig. 17** M1 thermal map for SALTUS Baseline Design

**Fig. 18** M1 temperature change in L2 environment

**Fig. 19** Concept for grounding features incorporated along gore seams (image credit L'Garde)

**Fig. 20** Angular flux density by the parent body source for micrometeoroids with momenta ≥ 1 μNs. Maps are in a solar-orbital coordinate frame centered on a spacecraft located at L1 where the center circle denotes the subsolar direction, P the prograde direction, and R the retrograde direction. Jupiter-Family Comets are expected to account for ~90% of the micrometeoroid flux in the range considered, with this population flux primarily concentrated in the sunward and antisunward directions. Figure originally from Thorpe et al. (2019) reproduced with permission.

**Fig. 21** Comparisons of the sky-averaged micrometeoroid integral flux above a particle mass threshold (m) predicted by the SPENVIS (solid blue line) and Thorpe et al. (2019) (solid red line). The models produce similar results for threshold masses > $10^{-8}$ g. The Thorpe et al. (2019) model is converted from a function of momenta to a function of mass assuming a particle velocity of 35 km/s. The dotted magenta lines show the bounds for particle velocities of v = 5 km/s and v = 70 km/s.

**Fig. 22** f(□) over SALTUS field of regard

**Fig. 23** Rate of micrometeoroid impacts above a given mass threshold, the isotropic Grun Flux (solid black trace), 2x Grun Flux (dash-dot blue trace) and 3x Grun Flux (dash-dot-dot green trace)

**Fig. 24** Kinetic energy of micrometeoroids impacting M1. The black line depicts the chosen velocity of 35 km/s for all particle masses. The dotted cyan and magenta lines show the approximate upper and lower bounds of kinetic energy for the full range of micrometeoroid velocities (~5-70 km/s).

**Fig. 25** Change in radius of M1 if all the kinetic energy of an impacting particle goes into compressing the gas. Particle masses above 7 x $10^{-6}$ g change the radius by more than 0.1%.

**Fig. 26** Effective spring constant, k, if compression of M1 is modeled as a spring.

**Fig. 27** Increase in pressure of M1 if the impacting particle's kinetic energy is assumed to go into compressing the gas. The red dotted line denotes the pressure tolerance of M1. This pressure tolerance is exceeded for particle masses greater than ~$10^{-5}$ g.

**Fig. 28** Fragmentation and propagation in gossamer structures, from [35]

**Fig. 29** Entrance and second layer damages from [36]

**Fig. 30** ½ mil VDA

**Fig. 31** ½ mil strengthened laminate

**Fig. 32** Contour plots of inflatant that must be vented to reach P0 when moving from the initial pointing to any other point in the FOR, the initial point for the far left panel is ±20° pitch, ±5° roll, for the middle panel, +15° pitch, 0,° roll and for the right most panel, 0° pitch, 0° roll

**Fig. 33** Map of Peak M1 Temperature